# Multi-Dimensional Theory of Protein Folding

Kazuhito Itoh and Masaki Sasai

Department of Computational Science and Engineering, Nagoya University, Nagoya, 464-8603, Japan

**ABSTRACT**

Theory of multi-dimensional representation of free energy surface of protein folding is developed by adopting structural order parameters of multiple regions in protein as multiple coordinates. Various scenarios of folding are classified in terms of cooperativity within individual regions and interactions among multiple regions and thus obtained classification is used to analyze the folding process of several example proteins. Ribosomal protein S6, src-SH3 domain, CheY, barnase, and BBL domain are analyzed with the two-dimensional representation by using a structure-based Hamiltonian model. Extension to the higher dimensional representation leads to the finer description of the folding process. Barnase, NtrC, and an ankyrin repeat protein are examined with the three-dimensional representation. The multi-dimensional representation allows us to directly address questions on folding pathways, intermediates, and transition states.

## I. INTRODUCTION

Elucidation of the mechanism of protein folding is a major challenge in molecular biological physics. Intense interest has been focused on how folding pathways are determined, how transition state ensembles are organized, and how these characteristics manifest in experiment (1-4). The energy landscape theory, which describes the statistical features of the free energy surface of protein conformational change, has provided a transparent view to these problems (3, 4). Since a protein conformation is determined by a vast number of atomistic degrees of freedom, the free energy surface of protein conformation is intrinsically high dimensional. The energy landscape theory provides a guideline to project this high dimensional surface onto a few reaction coordinates, which leads to a concise picture of folding (5). Especially, the one-dimensional representation of folding with a single reaction coordinate has been useful to distinguish between two-state and three or more state transitions (6) and to locate the transition state ensemble on the reaction coordinate.

In order to further analyze the folding mechanism, one should go beyond the one-dimensional representation by employing two or higher dimensional representation of the free energy surface. Merits of the multi-dimensional representation become evident when we examine competition among multiple parallel folding pathways which are not distinguishable in the one-dimensional representation (7). The multi-dimensional representation is powerful to analyze the heterogeneity of transition state ensemble and malleability of pathways (8-10). When multiple coordinates are chosen as order parameters of structural formation of individual domains of a protein, the multi-dimensional representation can depict the relative stability or instability of each domain and the cooperativity among them (9, 10). The coupled folding-binding process of protein complex should be analyzed by using coordinates for individual proteins and a coordinate for protein-protein interactions (11, 12). In this way, possibilities of the multi-dimensional representation have begun to be explored in folding study, but the systematic analysis of the representation scheme has not yet been fully developed.

In the early attempts of multi-dimensional representation, relation between the chain collapse and the formation of residue contacts was investigated (13, 14). Use of multiple coordinates which are similar to each other, as in the combined use of radius of gyration and the number of residue contacts, however, does not yield results much different from those derived from the one-dimensional representation. Instead, the use of *independent* multiple coordinates is needed to fully bring out the merits of the multi-dimensional representation. A straightforward way to choose independent coordinates is to use structural order parameters at different regions in protein. This representation has been used by Itoh and Sasai to study folding processes of a single-domain protein (8) and multi-domain proteins (11). In the present paper, construction of this multi-dimensional representation is systematically analyzed and its physical meaning is discussed with a phenomenological model. Furthermore, by using a structure-based Hamiltonian model, the method is applied to example proteins to demonstrate how the multi-dimensional representation resolves problems in folding.

In Section II, the one-dimensional free energy surface of a phenomenological model is explained as a basis for the multi-dimensional theory. In Sections III and IV, multi-dimensional representation is formulated and applied to a phenomenological model. For two-dimensional representation, two independent reaction coordinates are defined by the order parameter for the structure formation of the N-terminal half and the one for the C-terminal half of protein. Various scenarios of folding process are distinguished depending on the cooperativity within each half and interactions between two halves. In Section V, two-dimensional free energy surfaces of several example proteins, ribosomal protein S6 (15-19), src-SH3 domain (20-22), CheY (23-25), barnase (26-28), and BBL domain (29-37) are calculated by using a structure-based Hamiltonian model and the information obtained from the two-dimensional representation is compared with experiments. In Section VI, extension to three or higher dimensional representation is explained by using barnase, NtrC (38,39), and an ankyrin repeat protein (40-42) as example proteins. The last section is devoted to summary and discussions.

## II. ONE DIMENSIONAL PHENOMENOLOGICAL MODEL

### One-dimensional representation

Before going to the multi-dimensional theory, we discuss the one-dimensional representation of the free energy surface. Phenomenological models of one-dimensional surface have been studied by many authors, and the subtle balance between entropy and energy in these models has been discussed (43-50). We here revisit this issue to prepare for the multi-dimensional theory.

The reaction coordinate is defined by the order parameter of structure formation, $x = M/N$ with $0 \le x \le 1$, where $M$ is the number of residues which take the configuration similar to that in the native conformation and $N$ is the total number of residues in protein. The constrained partition function $Z(x)$ is

$$Z(x) = \mathrm{Tr}_x \exp(-\beta H) = \Omega(x) \big/ \langle \exp(\beta H) \rangle_x \,, \qquad (1)$$

where $H$ is Hamiltonian of the system, $\beta = 1/k_B T$, and $\mathrm{Tr}_x$ is a sum over the atomistic coordinates of the system under the constraint of the order parameter $x$. $\langle\cdots\rangle_x = \mathrm{Tr}_x[(\cdots)\exp(-\beta H)]/Z(x)$, and $\Omega(x) = \mathrm{Tr}_x[1]$. By writing

$$F_\mathrm{e}(x) = k_\mathrm{B}T\ln\langle\exp(\beta H)\rangle_x, \quad (2)$$

and

$$S_\mathrm{c}(x) = k_\mathrm{B}\ln\Omega(x), \quad (3)$$

free energy $F(x) = -k_\mathrm{B}T\ln Z(x)$ is

$$F(x) = F_\mathrm{e}(x) - TS_\mathrm{c}(x). \quad (4)$$

We call $F_\mathrm{e}(x)$ the effective energy. When we write $E(x) = \langle H\rangle_x$, $F_\mathrm{e}(x)$ can be decomposed as

$$F_\mathrm{e}(x) = E(x) - TS_\mathrm{e}(x), \quad (5)$$

with

$$-TS_\mathrm{e}(x) = k_\mathrm{B}T\ln\left\langle e^{\beta(H-E(x))}\right\rangle_x. \quad (6)$$

From the inequality $\langle e^{\beta H}\rangle_x \geq e^{\beta E(x)}$, we find $S_\mathrm{e}(x) \leq 0$. $S_\mathrm{c}(x)$ of Eq.3 is positive and takes account of all the available conformations under the constraint $x$. At a given temperature $T$, those conformations are not equally searched by the protein chain but are sampled with the Boltzmann weight, which leads to the effective reduction of the number of available states and this tendency is represented by the negative entropy $S_\mathrm{e}(x)$. $S_\mathrm{e}(x)$, therefore, is determined by the statistical features of the energy landscape. $-TS_\mathrm{e}(x)$ can be expanded as

$$-TS_\mathrm{e}(x) = \sum_{n=2}^{\infty}\frac{\beta^{n-1}}{n!}C_n(H;x), \quad (7)$$

where $C_n(H; x)$ is the $n$th order cumulant of $H$ under the constraint of $x$, e.g., $C_2(H; x) = \langle H^2\rangle_x - \langle H\rangle_x^2$ and $C_3(H; x) = \langle H^3\rangle_x - 3\langle H^2\rangle_x \langle H\rangle_x + 2\langle H\rangle_x^3$. Eq.7 shows that $S_\mathrm{e}(x)$ represents the multi-body correlation of interactions, so that $S_\mathrm{e}(x)$ determines the cooperativity in the folding process as explained in the following subsections.

**Cooperativity in folding transition**

To further proceed from the above general expression, we have to consider a more specific model. Here, we introduce a phenomenological model by defining the functional forms of $S_\mathrm{c}(x)$ and $F_\mathrm{e}(x)$. For $S_\mathrm{c}(x)$, it should be reasonable to assume the form used in Ref.44;

$$S_c(x) = k_B \ln\left\{\Omega(1)\nu^{N(1-x)}\binom{N}{Nx}\right\}, \tag{8}$$

where $\nu = [\Omega(0)/\Omega(1)]^{1/N}$ is the ratio of the phase-space volume that a residue having the nonnative configuration can take over the volume a residue having the native configuration can take, where we assume for simplicity that $\nu$ is independent of the residue position. The entropic cost for a residue to take the native configuration is, therefore, $k_B\ln\nu > 0$. Then, the functional form of $S_c(x)$ is determined by $N$ and $\nu$. $\partial S_c/\partial x$ is calculated as

$$\frac{\partial S_c(x)}{\partial x} = Nk_B\left[\psi(N(1-x)+1) - \psi(Nx+1) - \ln\nu\right], \tag{9}$$

where $\psi(x)$ is a polygamma function. $1/N\,\partial S_c/\partial x$ only slightly depends on $N$.

Here, we refer to pairs of interacting residues in the native conformation as native pairs and refer to interactions between native pairs in any conformation as native interactions. We assume $H$ as the effective Hamiltonian obtained by averaging solvent degrees of freedom. Then, $H$ can be decomposed to the part representing native interactions $H_n$, and the rest part including nonnative interactions $H_{nn}$, as $H = H_n + H_{nn}$. It should be reasonable to choose $H_n$ to be a Go-type Hamiltonian. The effective energy $F_e(x)$ can be decomposed as

$$F_e(x) = k_BT\ln\langle\exp(\beta H_n)\rangle_x^{H_n} + k_BT\ln\langle\exp(\beta H_{nn})\rangle_x, \tag{10}$$

where $\langle\cdots\rangle_x^{H_n}$ is the average taken with Hamiltonian $H_n$ under the constraint of $x$.

$\ln\langle\exp(\beta H_n)\rangle_{x=0}^{H_n} \approx 0$, and $\ln\langle\exp(\beta H_{nn})\rangle_{x=1} = 0$.

$F_e(x)$ can be written as $F_e(x) = -[F_e(0) - F_e(1)]g(x) + F_e(0)$ with $g(0) = 0$ and $g(1) = 1$. Writing the total number of native pairs as $n_{tot}$, using $\varepsilon = [F_e(0) - F_e(1)]/n_{tot}$, and putting $F_e(0) = 0$ without loss of generality, we have

$$F_e(x) = -\varepsilon n_{tot}\, g(x), \tag{11}$$

Since $F(0) - F(1) = F_e(0) - F_e(1) - Nk_BT\ln\nu$, $F_e(0) - F_e(1) > 0$ is a necessary condition to be $F(0) > F(1)$, and hence $F_e(0) - F_e(1) > 0$ is a necessary condition to draw a reasonable free energy surface which allows the folding/unfolding transition. We, therefore, consider the situation of $\varepsilon > 0$ in Eq.11.

In many natural proteins, the funnel-like feature should be well developed in their folding energy landscapes, so that $k_BT\ln\langle\exp(\beta H_{nn})\rangle_x$ in Eq.10, which involves frustrated interactions and brings about ruggedness in energy landscapes, should only make minor contributions and $k_B T \ln\langle \exp(\beta H_n) \rangle_x^{H_n}$ dominates the overall change of the value of $F_e(x)$. Since $\langle H_n \rangle_x^{H_n}$ decreases from $x = 0$ to $x = 1$ by gaining native interactions, it is reasonable to assume as the first approximation that $F_e(x)$ is a decreasing function of $x$. From the expression of $F_e(x) = -\varepsilon n_{tot} g(x)$, we can see that $g(x)$ should be a monotonously increasing function of $x$ from $g(0) = 0$ to $g(1) = 1$. Then, the curvature of $g(x)$ determines the primary features of $F_e(x)$ and hence the free energy surface $F(x)$. We highlight this point by assuming $g(x) = x^\alpha$. We then have the form,

$$F_e(x) = -\varepsilon n_{tot} x^\alpha . \tag{12}$$

In this way, cooperativity of the folding process is expressed by the exponent $\alpha$ in Eq.12. Using Eqs.4, 8, and 12, effects of cooperativity can be examined by calculating $F(x)$ and $\partial F(x)/\partial x = \partial F_e(x)/\partial x - T\partial S_c(x)/\partial x$ with different values of $\alpha$. We can see in Fig.1a and Fig.1b that $F(x)$ has two minima for $\alpha > \alpha_c$ corresponding to the unfolded and folded states and the protein chain undergoes the two-state transition between them. The barrier in $F(x)$, which separates the unfolded and folded states, becomes higher as $\alpha$ becomes larger. At around $\alpha \approx \alpha_c$, the transition between two states is a process passing through a negligibly small or no barrier, which can be regarded as downhill folding. For $\alpha < \alpha_c$, $F(x)$ has a single minimum and folding is a non-cooperative process. As shown in Fig.1c, $\alpha_c$ decreases as $\ln \nu$ increases, e.g., $\alpha_c = 2.85$ for $\ln \nu = 1.0$, and $\alpha_c = 2.39$ for $\ln \nu = 1.5$ at $T = T_F$. Features of the transition are clearly seen when we plot $F(x)$ and the population of conformations $P(x) \propto \exp(-F(x)/k_B T)$ at different temperatures. Temperature dependence of $F(x)$ and $P(x)$ is shown in Figs.1d, 1e, and 1f for different values of $\alpha$ by assuming that the temperature dependence of $\alpha$ can be neglected in the temperature range examined in Fig.1.

When $k_B T \ln\langle \exp(\beta H_{nn}) \rangle_x$ in Eq.10 is negligible and the system is described by the Hamiltonian $H_n$, the expression of Eq.12 is indeed validated: The average energy in the Hamiltonian should be $E(x) \approx -\varepsilon n(x)$, where $-\varepsilon$ is the energy decrease to form a native interaction, and $n(x)$ is the average number of native interactions in a conformation of $x$. Then, $F_e(x)$ can be written as $F_e(x) = -\varepsilon n_{tot} g(x)$, where $g(x) \approx n(x)/n_{tot} + T/(\varepsilon n_{tot}) S_e(x)$. $n(x)$ should be an increasing function of $x$ satisfying $n(0) = 0$ and $n(1) = n_{tot}$. Hamiltonian-dependent part of entropy should be $S_e(0) \approx 0$, $S_e(1) \approx 0$, and $S_e(0) < 0$ for $0 < x < 1$, so that $S_e(x)$ is an intrinsically nonlinear function of $x$. When $n(x)$ dominates the overall behavior of $g(x)$, $g(x)$ is a monotonously increasing function of $x$, but its curvature or the exponent $\alpha$ should be strongly dependent on the nonlinearity in $S_e(x)$. As will be explicitly shown in Section IV, assumption of Eq.12 is justified when we use a Go-like structure-based Hamiltonian.

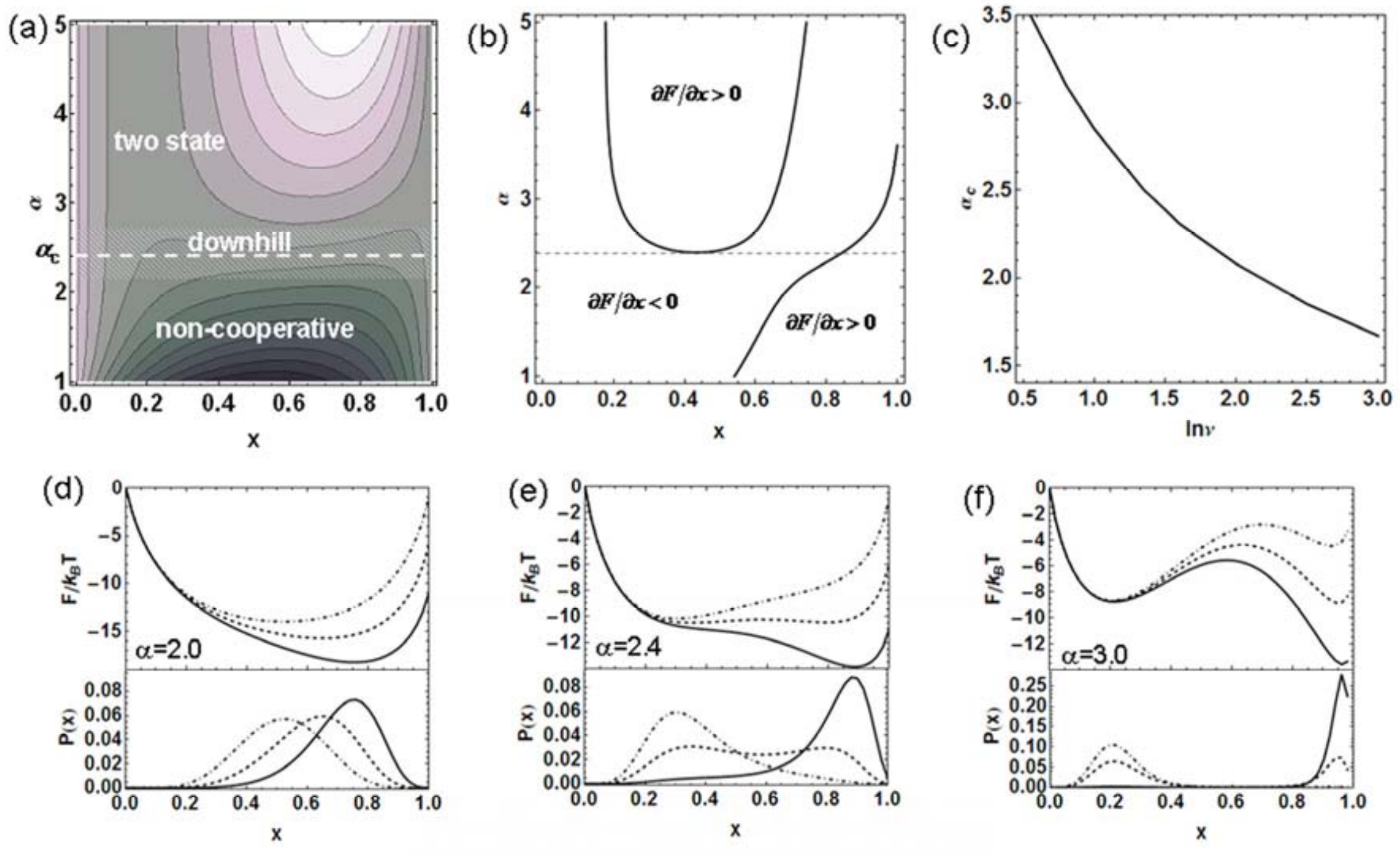

**FIGURE 1**. One-dimensional representation of free energy surface of the phenomenological model. (a) $F(x)$, as a function of $x$ and $\alpha$. Contour lines are drawn in every $3k_BT$. (b) Lines of $\partial F(x)/\partial x=0$ with $\ln\nu = 1.5$ at the folding temperature. (c) $\alpha_c$ as a function of $\ln\nu$ at the folding temperature. Shown are one-dimensional free energy surfaces $F(x)$ and populations $P(x)$ at different temperatures with $\ln\nu = 1.5$ at $\alpha = 2.0$ (d), $\alpha = 2.4$ (e), and $\alpha = 3.0$ (f). In (a-f) $N$=50.

**Nonnative interactions may affect the cooperativity**

We should note that the ruggedness of the energy landscape arising from the nonnative interactions in the second term of Eq.10 affects the functional form of $F_e(x)$ by modifying $g(x)$ as $g(x) = x^{\alpha} + g_{rugg}(x)$. The effective value of $\alpha$ can be modified due to this contribution. When $\varepsilon n_{tot}|g_{rugg}(x)|$ is smaller than the height of the free energy barrier, however, this modification does not affect the main features of folding. In the case that the cumulant of $H_{nn}$ in $k_BT\ln\langle\exp(\beta H_{nn})\rangle_x$, can be truncated at the 2nd order, the ruggedness gives rise to expressions of free energy $F_{rugg}(x,T) = -\Delta E^2(x,T)/2k_BT$ and energy $E_{rugg}(x,T) = -\Delta E^2(x,T)/k_BT+(1/2k_B)\partial[\Delta E^2(x,T)]/\partial T$, where $\Delta E^2(x,T)$ is the 2nd order cumulant, which represents the contribution of the ruggedness, and the second term of $E_{\rm rugg}(x,T)$ is the contribution from correlations between $H_{\rm n}$ and $H_{\rm nn}$ (Derivation is explained in APPENDIX A). When $\Delta E^2$ is a temperature-independent function, this expression is reduced to the result ignoring the correlations (44). Furthermore, the folding/unfolding rates become smaller due to the "friction" effects of ruggedness (45). Hereafter in this paper, we do not consider these effects of ruggedness but focus on the global features of the free energy surface which determines the folding pathway.

**Cooperativity is determined by the nonlinearity in $S_{\rm e}(x)$**

As shown in Eq.7, the nonlinearity in $S_{\rm e}(x)$ is determined by the correlation among interactions. The chain connectivity or multi-body interactions should be the origin of such correlation. Importance of $S_{\rm e}(x)$ to determine the curvature of free energy surface and hence to determine the cooperativity of folding is clarified by explicitly calculating energy $E(x)$ and the Hamiltonian-dependent part of entropy $S_{\rm e}(x)$ from Eq.11:

$$E(x) = \frac{\partial(\beta F(x))}{\partial\beta}$$

$$= -\varepsilon n_{\rm tot} g(x) - \beta n_{\rm tot}\frac{\partial(\varepsilon g(x))}{\partial\beta}, \qquad (13)$$

and

$$S_{\rm e}(x) = -\, k_{\rm B}\beta^2 n_{\rm tot}\frac{\partial(\varepsilon g(x))}{\partial\beta}. \qquad (14)$$

Since $\partial(g(1))/\partial\beta = \partial(g(0))/\partial\beta = 0$, Eqs. 13 and 14 satisfy $E(1) = -[\varepsilon - T(\partial\varepsilon/\partial T)]n_{\rm tot}$ , $E(0) = 0$, $S_{\rm e}(1) = (\partial\varepsilon/\partial T)n_{\rm tot}$, and $S_{\rm e}(0) = 0$. It should be reasonable to assume that the temperature dependence of $\varepsilon$ is small. When the temperature dependence of $\varepsilon$ is negligible, $E(1) = -\varepsilon n_{\rm tot}$ and $S_{\rm e}(1) = 0$. By assuming Eq.12 and neglecting the temperature dependence of $\varepsilon$, $E(x)$ and $S_{\rm e}(x)$ are given by

$$E(x) = -\varepsilon n_{\rm tot}[1 + \beta\frac{\partial\alpha}{\partial\beta}\ln x]x^{\alpha}, \qquad (15)$$

and

$$S_{\rm e}(x) = -\varepsilon n_{\rm tot}\beta^2\frac{\partial\alpha}{\partial\beta}k_{\rm B}x^{\alpha}\ln x. \qquad (16)$$

Since $S_{\rm e}(x) \le 0$ and $x^{\alpha}\ln x < 0$ for $0 < x < 1$, $\partial\alpha/\partial\beta$ must be negative. See Fig.2 for the functional form of $S_{\rm e}(x)$. From Eqs. 15 and 16, we can see that the larger $|\partial\alpha/\partial\beta|$ is, the larger $x$ dependence $S_{\rm e}(x)$ has and the smaller $x$ dependence $E(x)$ has. This point is clarified by drawing $F_{\rm e}(x)$ of Eq.12 and $E(x)$ of Eq.15 in Fig.3a: Addition of $-TS_{\rm e}(x)$ to $E(x)$ enhances the dependence of $F_{\rm e}(x)$ on $x$, leading to the larger curvature of $F_{\rm e}(x)$ than that of $E(x)$, which brings about the higher barrier separating the unfolded and folded states in $F(x)$ as shown in Fig.3b. The role of $-TS_e(x)$ can be further confirmed by fitting $E(x)$ as $E(x) = -\varepsilon n_{tot} x^{\alpha'}$. When $|\beta(\partial\alpha/\partial\beta)| \ll 1$, we can see from Eq.15 that exponents are related by $\alpha \approx \alpha' - \beta(\partial\alpha/\partial\beta)$, so that $\alpha$ and hence the cooperativity in folding process increases as the contribution of $\partial\alpha/\partial\beta < 0$ is larger, which shows that the larger dependence of $S_e(x)$ on $x$ leads to the larger cooperativity. This role of $-TS_e(x)$ to enhance the cooperativity is schematically illustrated in Fig.4. As $S_e(x)$ should depend on the strength of multi-body interactions, $\alpha$ here should also depend on multi-body interactions as pointed out in the earlier landscape theory by Plotkin et al. (50).

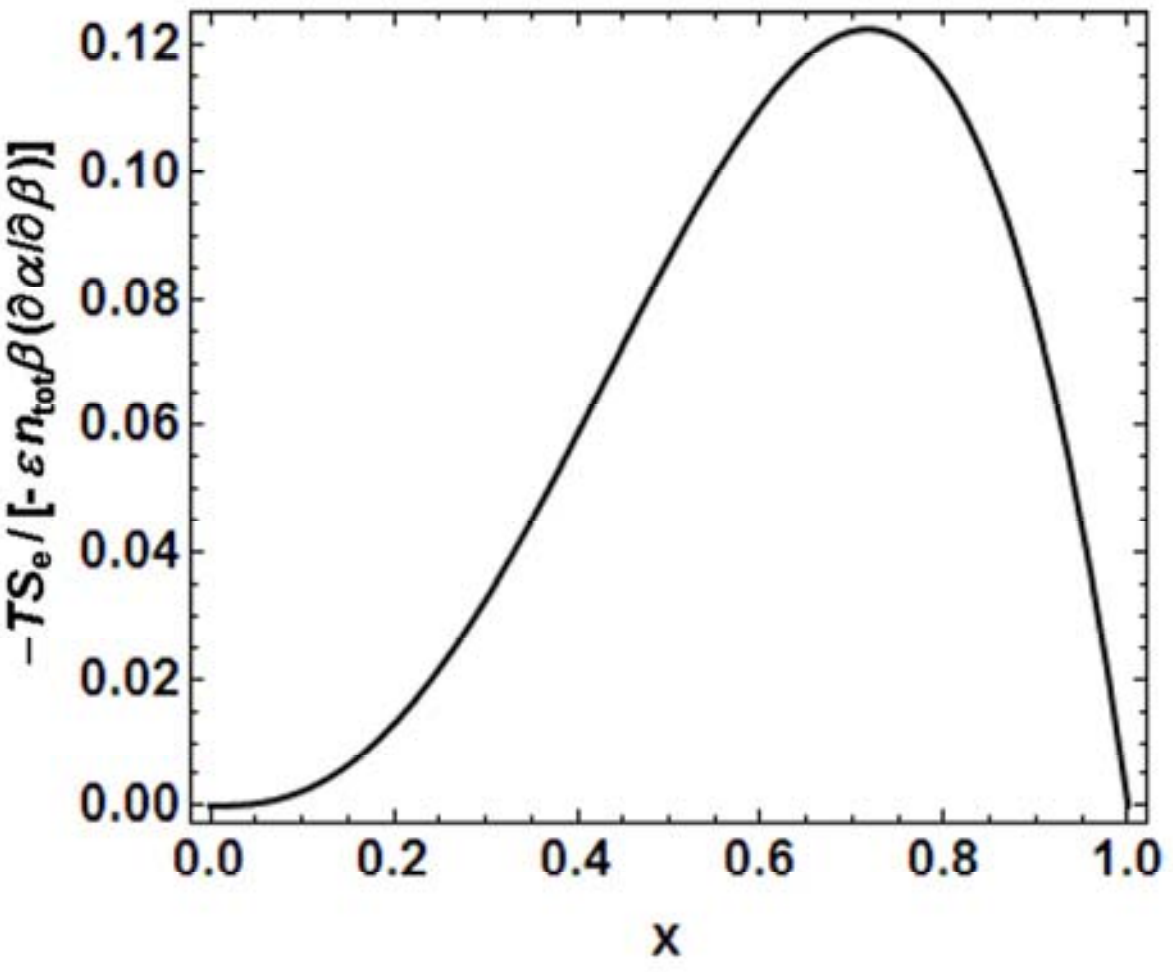

**FIGURE 2** Functional form of $-TS_e(x)$ of the one-dimensional phenomenological model. Shown is $S_e(x)$ of Eq.14 normalized by the factor $-T/[-\varepsilon n_{tot}\beta(\partial\alpha/\partial\beta)]$. $-TS_e(x)$ has a large value in $0 < x < 1$ though $S_e(0) = S_e(1) = 0$.

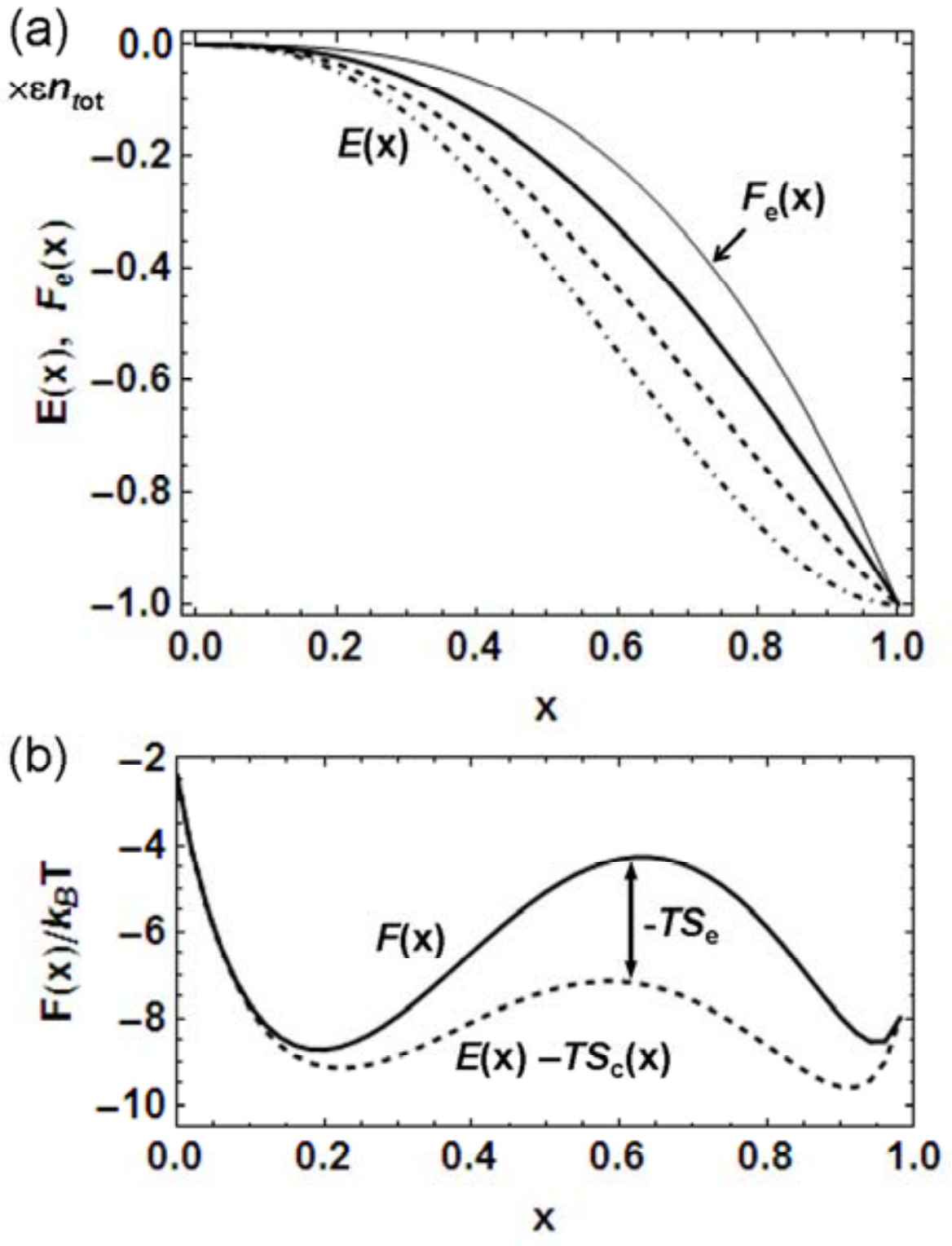

**FIGURE 3** (a) The energy function $E(x)$ of Eq.13 with $\beta(\partial\alpha/\partial\beta) = -1.0$ (thick real line), $-2.0$ (dashed line), and $-3.0$ (dot-dashed line), and the function $F_e(x)$ (thin real line) of the one-dimensional phenomenological model. $F_e(x)$ has the larger curvature than $E(x)$ owing to the contribution of $\beta(\partial\alpha/\partial\beta)$. (b) Comparison between one-dimensional free energy surface $F(x) = E(x) - T[S_e(x) + S_c(x)]$ (real line) and $E(x) - TS_c(x)$ (dashed line) with $\ln\nu=1.5$, $\alpha = 3.0$ and $\beta(\partial\alpha/\partial\beta) = -0.2$ at the folding temperature. Addition of $-TS_e(x)$ to $E(x)$ enhances cooperativity in folding/unfolding.

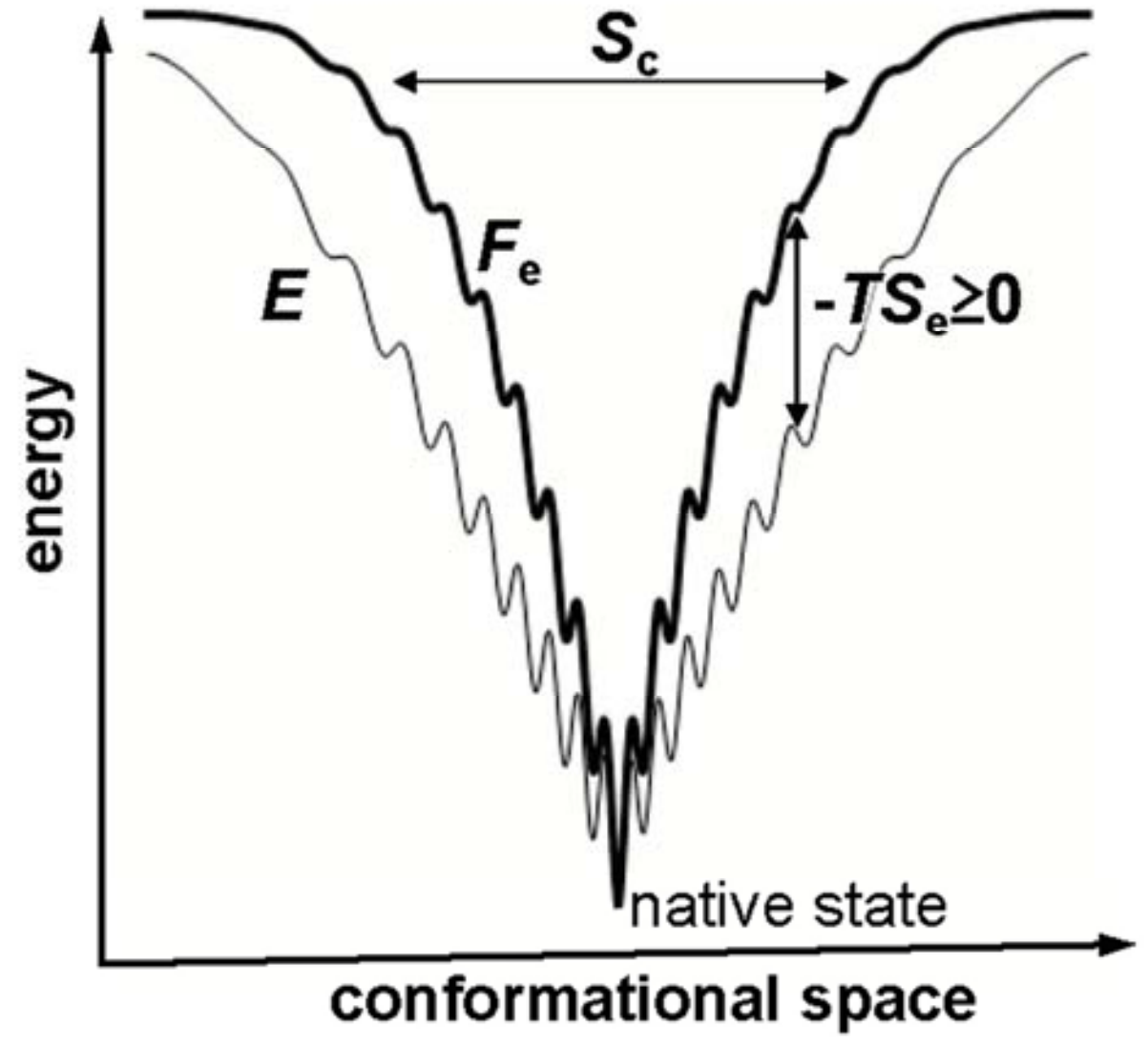

**FIGURE 4** Schematic illustration of the *effective energy landscape*. While $E = \langle H \rangle$ has a global tendency to decrease as the protein chain approaches the native state, the functional form of the effective energy, $F_e = E - TS_e$ , is largely determined by the entropic factor, $-TS_e \geq 0$, which represents the higher order correlations of $H$ through the terms of $\langle H^n \rangle$ ($n \geq 2$). The width of the effective funnel represents the conformational entropy $S_c$.

More complex folding process such as the three-state folding transition can be expressed by adopting a more complex form of $F_e(x)$ than Eq.12, but we show later in Sections IV and V that such process can be more appropriately described by the multi-dimensional representation.

## III. MULTI-DIMENSIONAL REPRESENTATION

In this section we explain construction of multi-dimensional representation of free energy surface by extending the discussion of the last section. A straightforward way to define multiple independent coordinates is to use order parameters of structure formation of independent regions in a protein. When the protein chain is divided into $L$ regions, $L$ order parameters are $x_i = M_i / N_i$ with $i$ = 1, 2, …, $L$, where $N_i$ is the number of residues in the $i$th region and $M_i$ is the number of residues taking the native-like configuration in the $i$th region. $N = \sum_{i=1}^{L} N_i$ and $M = \sum_{i=1}^{L} M_i$ . Then, the $L$ reaction coordinates are defined by $\mathbf{x} = (x_1, x_2, \cdots, x_L)$ .

By extending the expression of Eq.3, conformational entropy $S_c$ should be represented by

$$S_c(\mathbf{x}) = \sum_{i=1}^{L} S_c^{(i)}(x_i) + k_B \ln \Omega(\mathbf{1}), \qquad (17)$$

with $\Omega(\mathbf{1}) = \mathrm{Tr}_{\mathbf{x}=\mathbf{1}}[1]$. We can use, for example, the functional form of

$$S_{\rm c}^{(i)}(x_i) = k_{\rm B}\ln\left\{\nu^{N_i(1-x_i)}\binom{N_i}{N_i x_i}\right\}. \tag{18}$$

Free energy in the $L$-dimensional space is written as $F(\mathbf{x}) = F_{\rm e}(\mathbf{x}) - TS_{\rm c}(\mathbf{x})$ with

$$\begin{aligned} F_{\rm e}(\mathbf{x}) &= E(\mathbf{x}) - TS_{\rm e}(\mathbf{x}) \\ &= k_{\rm B}T\ln\langle \exp(\beta H)\rangle_{\mathbf{x}} \end{aligned}, \tag{19}$$

where energy is $E(\mathbf{x}) = \langle H\rangle_{\mathbf{x}}$, the Hamiltonian-dependent part of entropy is $S_{\rm e}(\mathbf{x}) = -k_{\rm B}\ln\langle\exp[\beta(H-E(\mathbf{x}))]\rangle_{\mathbf{x}}$, and $\langle\cdots\rangle_{\mathbf{x}}$ is the average taking with Hamiltonian $H$ under the constraint of $\mathbf{x}$.

Hamiltonian can be decomposed as

$$H = \sum_i h_i + \sum_{(i,j)} h_{ij} + \sum_{(i,j,k)} h_{ijk} + \cdots, \tag{20}$$

where $h_i$ represents interactions between residues within the region $i$, $h_{ij}$ represents interactions between residues in the $i$th region and residues in the $j$th region, and $h_{ijk}$ is the term representing three or more-body interactions among residues belonging to $i$, $j$, and $k$th regions. Summing up the multi-region terms as

$$V = \sum_{(i,j)} h_{ij} + \sum_{(i,j,k)} h_{ijk} + \cdots,$$

the effective energy $F_{\rm e}(\mathbf{x})$ is

$$F_{\rm e}(\mathbf{x}) = \sum_{i=1}^{L} F_{\rm e}^{(i)}(x_i) + U(\mathbf{x}), \tag{21}$$

where $F_{\rm e}^{(i)}(x_i)$ is the intra-region free energy obtained by subtracting $-TS_{\rm c}^{(i)}(x_i)$, to which we should refer as the intra-region effective energy;

$$F_{\rm e}^{(i)}(x_i) = k_{\rm B}T\ln\langle\exp(\beta h_i)\rangle_{x_i}^{h_i}, \tag{22}$$

and $U(\mathbf{x})$ is the inter-region free energy;

$$U(\mathbf{x}) = k_{\rm B}T\ln\langle\exp(\beta V)\rangle_{\mathbf{x}}. \tag{23}$$

Here, $\langle\cdots\rangle_{x_i}^{h_i}$ is the average taken with the $i$th part of Hamiltonian $h_i$ under the constraint of $x_i$. As in the case of one-dimensional representation, $F_{\rm e}^{(i)}(x_i)$ of Eq.22 can be decomposed into energy and entropy terms as $F_{\rm e}^{(i)}(x_i) = E_0^{(i)}(x_i) - TS_{\rm e0}^{(i)}(x_i)$ with $E_0^{(i)}(x_i) = \langle h_i \rangle_{x_i}^{h_i}$, and $-TS_{\mathrm{e}0}^{(\mathrm{i})}(x_i) = k_B T \left\langle \exp[\beta(h_i - \langle h_i \rangle_{x_i}^{h_i})] \right\rangle_{x_i}^{h_i}$. In the similar way, $U(\mathbf{x})$ of Eq.23 can be decomposed as $U(\mathbf{x}) = E_U(\mathbf{x}) - TS_{\mathrm{e}U}(\mathbf{x})$ with $E_U(\mathbf{x}) = \langle V \rangle_{\mathbf{x}}$, and $-TS_{\mathrm{e}U}(\mathbf{x}) = k_\mathrm{B}T\ln\langle \exp[\beta(V - \langle V \rangle_{\mathbf{x}})] \rangle_{\mathbf{x}}$. Although $F_\mathrm{e}(\mathbf{x}) = E(\mathbf{x}) - TS_\mathrm{e}(\mathbf{x})$ can be decomposed into the single-region parts and the multi-region part as in Eq.21, we should note that either $E(\mathbf{x})$ or $S_\mathrm{e}(\mathbf{x})$ cannot be decomposed in such a way but is written with the residual term as

$$E(\mathbf{x}) - \sum_i E_0^{(i)}(\mathbf{x}) - E_U(\mathbf{x}) = \left\langle \left[ 1 - \frac{\exp(\beta V)}{\langle \exp(\beta V) \rangle_{\mathbf{x}}} \right] \sum_i h_i \right\rangle_{\mathbf{x}} . \qquad (24)$$

Using $F_\mathrm{e}^{(i)}(x_i)$, the total free energy is written by a sum of the single-region part $F_0(\mathbf{x})$ and the multi-region part $U(\mathbf{x})$ as

$$F(\mathbf{x}) = F_0(\mathbf{x}) + U(\mathbf{x}), \qquad (25)$$

where $F_0(\mathbf{x}) = \sum_{i=1}^{L} F_{0i}(x_i)$ and

$$F_{0i}(x_i) = F_\mathrm{e}^{(i)}(x_i) - TS_\mathrm{c}^{(i)}(x_i). \qquad (26)$$

Since $F_{0i}(x_i)$ is the one-dimensional representation of each region, we can follow the discussion of Section II for $F_{0i}(x_i)$: The functional form of $F_\mathrm{e}^{(i)}(x_i)$ largely determines $F_{0i}(x_i)$, so that the functional form of $F_\mathrm{e}^{(i)}(x_i)$ determines whether each region tends to fold in a two-state or a downhill manner, where the functional form of $F_\mathrm{e}^{(i)}(x_i)$ is largely affected by the nonlinearity in $S_{\mathrm{e}0}^{(\mathrm{i})}(x_i)$. On the other hand, when $U(\mathbf{x})$ has a weak dependence on $\mathbf{x}$, each region tends to fold independently and when $U(\mathbf{x})$ has a strong dependence on $\mathbf{x}$, the multiple regions tend to fold cooperatively. In this way, we can analyze the hierarchical structure of cooperativity in folding by the combined use of $F_\mathrm{e}^{(i)}(x_i)$ and $U(\mathbf{x})$: Cooperativity in each region can be analyzed by using $F_\mathrm{e}^{(i)}(x_i)$ and cooperativity among multiple regions can be analyzed through $U(\mathbf{x})$. In Sections IV and V, we discuss the most straightforward way of division of protein into regions of $N_1 \approx N_2$ for $L = 2$. In Section VI, we show that further insights into the free energy surface and pathways can be gained by examining various different ways of division.

## IV. TWO-DIMENSIONAL PHENOMENOLOGICAL MODEL

In this section, we apply the method developed in Section III to a phenomenological model to illustrate how the multi-dimensional representation helps to parse the folding mechanisms. Various types of folding are classified with a phenomenological model, which should offer a language to describe the more complex folding processes, so that the results of this section will be used in Section V to distinguish folding schemes of example proteins.

We choose two regions, one with $N_1$ residues and the other with $N_2$ residues with $N = N_1 + N_2$. For two-domain proteins, it is natural to define each of two domains as each region (10), but for single-domain proteins, it should be convenient to use two regions of almost equivalent size. Though other ways of division of protein are possible, we here focus on the almost equal division of $N_1 \approx N_2 \approx N/2$ to demonstrate how the folding scenarios are classified with the multi-dimensional representation.

For $L = 2$, free energy is decomposed as

$$F(x_1,x_2) = F_{01}(x_1) + F_{02}(x_2) + U(x_1,x_2). \quad (27)$$

In the present phenomenological model, the expression of Eq. 18 is used for $S_{\mathrm{c}}^{(1)}(x_1)$ and $S_{\mathrm{c}}^{(2)}(x_2)$. We adopt the function of Eq. 12 as the intra–region effective energy functions for $i$ = 1 and 2 as

$$F_{\mathrm{e}}^{(\mathrm{i})}(x_i) = -\varepsilon n_{\mathrm{i}} x_i^{\alpha_i}, \quad (28)$$

and assume the inter-region free energy function to be

$$U(x_1,x_2) = -\varepsilon n_{12} x_1^{\gamma_1} x_2^{\gamma_2}, \quad (29)$$

where $n_i$ is the total number of native pairs between residues in the $i$th region, and $n_{12}$ is the total number of native pairs between residue in the 1st region and residue in the 2nd region. The total number of native pairs in the protein is $n_{\mathrm{tot}} = n_1 + n_2 + n_{12}$. From Eqs.28 and 29, we can see that the larger $\alpha_i$ is, the larger cooperativity each domain exhibits and the larger $n_{12}/n_{\mathrm{tot}}$ is, the two regions behave more cooperatively. We discuss the folding schemes of this model in several different cases.

**Case i**

We first consider the case that the protein is symmetrical and two regions are almost same as $n_1 = n_2$, $\alpha_1 = \alpha_2 = \alpha$, and $\gamma_1 = \gamma_2$. This case might be realized when the protein consists of two identical domains. In this idealized case, the folding behaviors are classified into four types as shown in Fig.5: (type I) The free energy surface has a single minimum and the protein exhibits non-cooperative folding like a helix-coil transition or downhill folding. (type II) The free energy surface has two minima corresponding to the unfolded and folded states, so that the folding is a two-state transition. A single saddle separates the unfolded and folded states. (type III) The folding is a two-state transition but the free energy surface has two saddles, so that there are two pathways passing through either of two saddles. (type IV) There are four minima in the free energy surface, two of which corresponds to the unfolded and folded states and the other two are kinetic intermediate states. There are two parallel pathways passing through either of these two intermediate states.

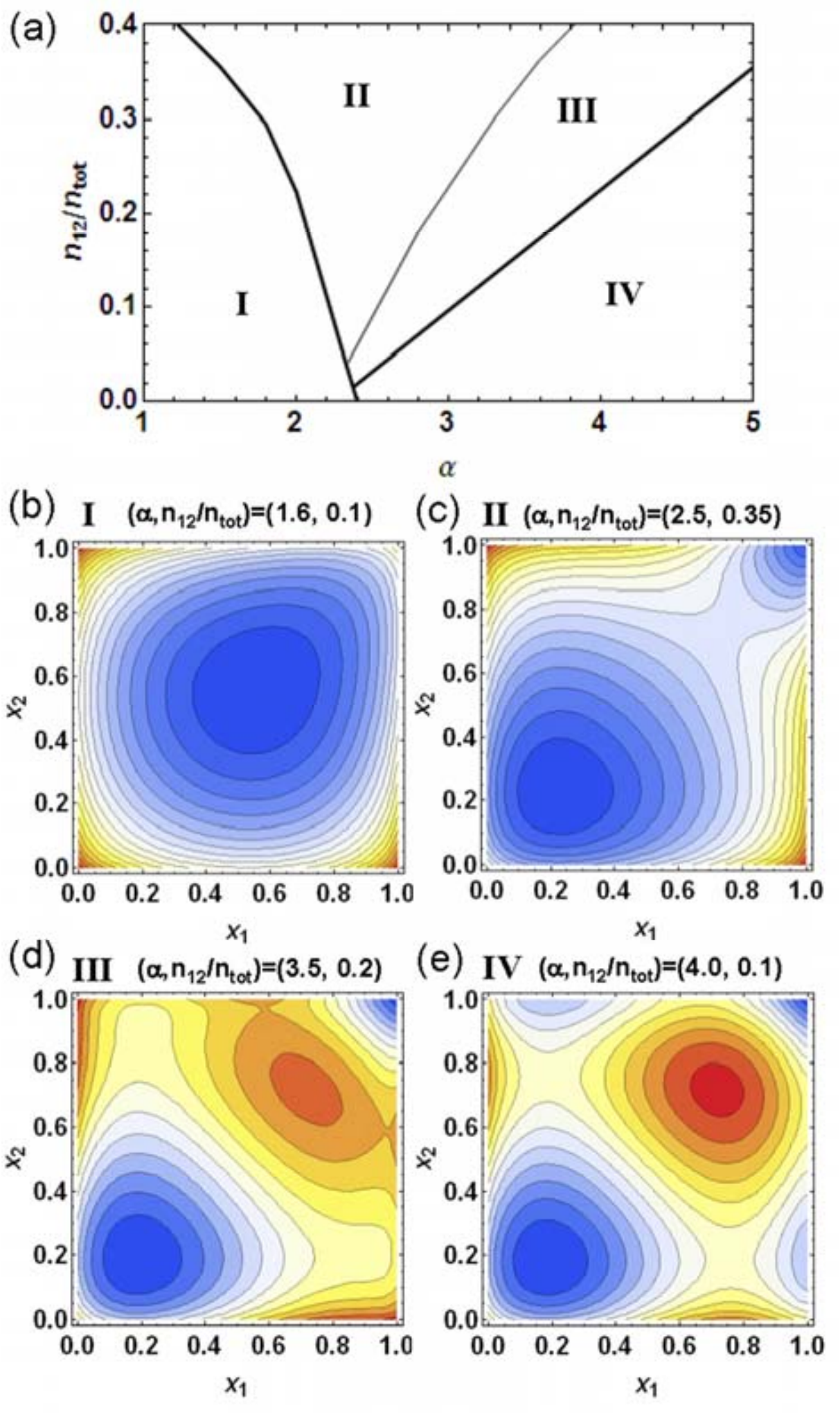

**FIGURE 5** Folding process of case i classified by the two-dimensional representation of free energy surface. Shown are phase diagram (a), free energy surface $F(x_1,x_2)$ for type I (b), type II (c), type III (d), and type IV (e). Contour lines are drawn in every $2k_BT$.

When there are two parallel pathways (type III and IV) these two pathways have equal statistical weights because the free energy surface is symmetrical. Individual protein molecules, however, take either of two pathways. This is the *symmetry-breaking* at the molecular level (7) and could be detected by single-molecule measurement. Along one pathway, the structure of region 1 is formed first, which catalyzes formation of region 2, while along the other pathway, region 2 is formed first, which catalyzes formation of region 1

As shown in Fig.5a, these four types appear depending on the values of $n_{12}/n_{\mathrm{tot}}$ and $\alpha$. Type I appears when $\alpha$ is small while type IV appears when $\alpha$ is large. Type II or Type III is realized when $n_{12}/n_{\mathrm{tot}}$ is large.

**Case ii**

In more generic cases, we cannot expect the idealized symmetry as in case i. In case ii, we treat the situation of $n_1 \neq n_2$, $\alpha_1 = \alpha_2 = \alpha$, and $\gamma_1 = \gamma_2$. The folding behaviors are classified into type I, type II, and type III as shown in Fig.6: (type I) The free energy surface has a single minimum and the protein exhibits non-cooperative or downhill folding. (type II) The free energy surface has two minima and the folding proceeds as two-state transition along a single folding pathway. Type II is further classified into two subtypes: In type IIa, two regions fold simultaneously along the folding pathway, while in type IIb, the region with larger $n_i$ folds faster and catalyzes the structure formation of the region of smaller $n_i$. (type III) The folding proceeds by passing through an intermediate and the protein exhibits three-state transition. At the intermediate, the region of larger $n_i$ is structured but the rest part is unfolded. The structure formation of the region of larger $n_i$ catalyzes folding of the rest part. In type IIIa. the intermediate is not the lowest minimum in free energy and appears as a kinetic intermediate. In type IIIb, the intermediate can become the lowest minimum i.e. the thermodynamically most stable state depending on the value of $\varepsilon$ or $k_B T$.

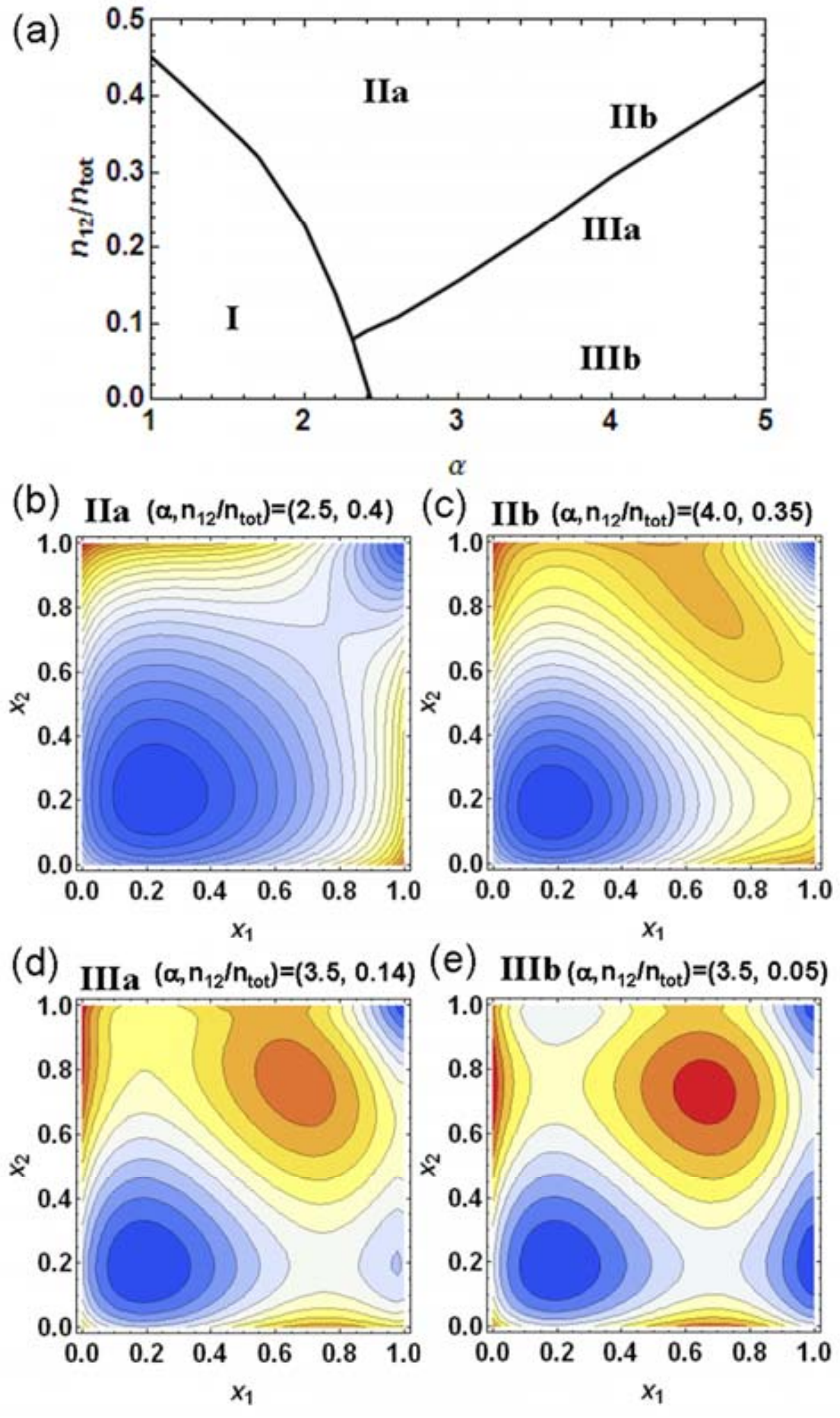

**FIGURE 6** Folding process of case ii classified by the two-dimensional representation of free energy surface. (a) Phase diagram with $(n_1 - n_2)/n_{\mathrm{tot}} = 0.05$ and $\gamma_1 = \gamma_2 = 3$, and shown are free energy surface $F(x_1,x_2)$ for type IIa (b), type IIb (c), type IIIa (d), and type IIIb (e). Free energy surface for type I of case ii resembles Fig. 6b. Contour lines are drawn in every $2k_{\mathrm{B}}T$.

As shown in Fig.6a, Type I is realized when $\alpha$ is small, while type III is realized when $\alpha$ is large. Type II appears when $n_{12}/n_{\mathrm{tot}}$ is large. Varying $\alpha_i$ or $n_{12}/n_{\mathrm{tot}}$ by mutation or changing environmental conditions, the folding behavior can be transformed between the three-state transition of type IIIa and the two-state transition of type IIb.

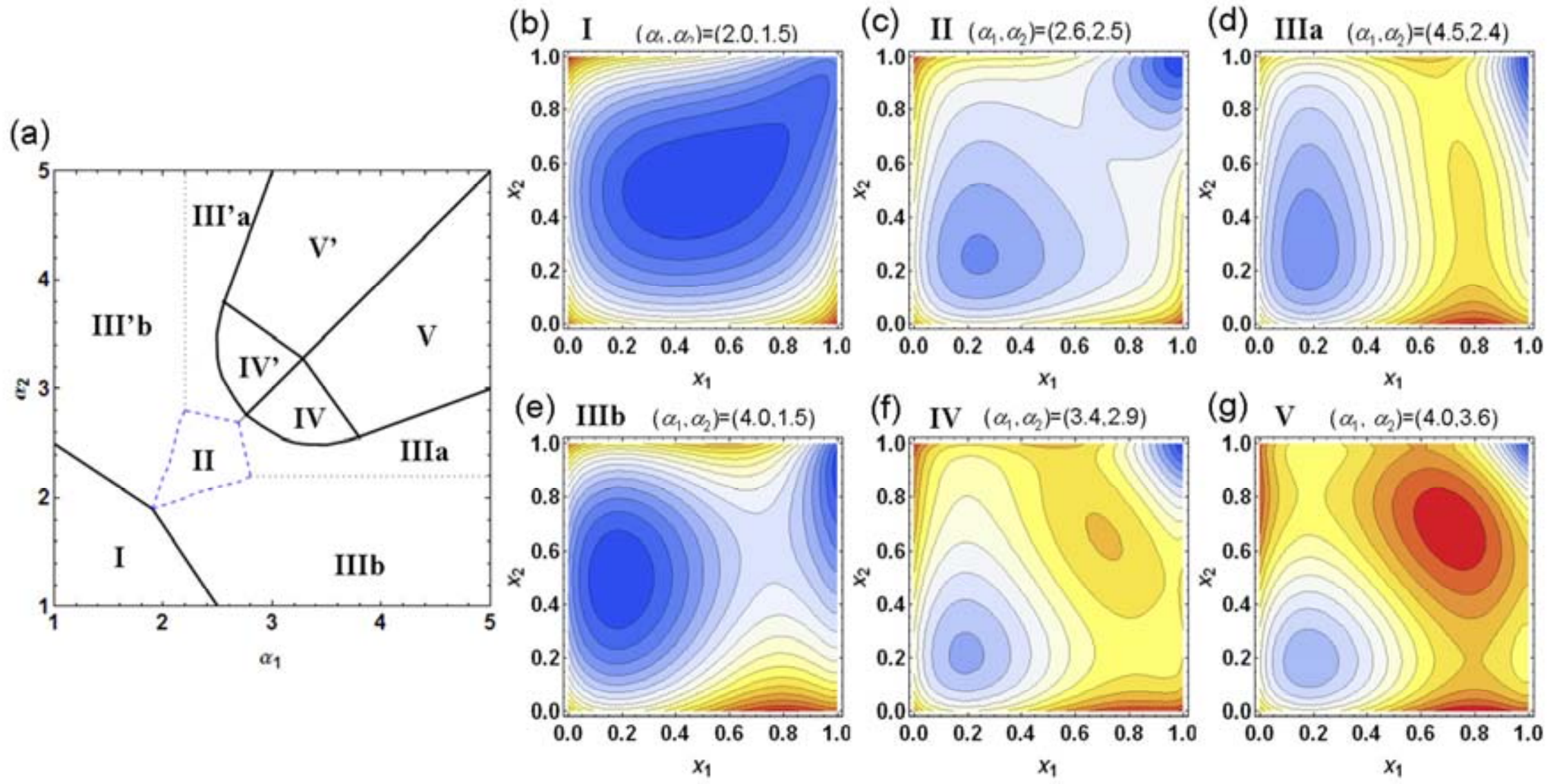

**FIGURE 7** Folding process of case iii classified by the two-dimensional representation of free energy surface. (a) Phase diagram with $n_{12}/n_{\mathrm{tot}} = 0.2$ and $\gamma_1 = \gamma_2 = 3$, and shown are free energy surface $F(x_1,x_2)$ for type I (b), type II (c), type IIIa (d), type IIIb (e), type IV (f), and type V (g). Contour lines are drawn in every $2k_B T$.

**Case iii**

We consider the case of $n_1 = n_2$, $\alpha_1 \neq \alpha_2$, and $\gamma_1 = \gamma_2$. There are five different types as shown in Fig.7: (type I) The free energy has a single minimum and the protein shows non-cooperative or downhill folding. (type II) The two-state transition proceeds almost symmetrically with simultaneous folding of two regions. (type III) The free energy is shallower for the structural change of the region of smaller $\alpha_i$, so that the region of smaller $\alpha_i$ is structurally formed first as in downhill folding and then the rest part folds. In type IIIa, the difference between two regions is distinct and in type IIIb, the difference is milder. The basin of folded state is also shallow, leading to the large fluctuation of the region of smaller $\alpha_i$ in the native state. (type IV) The two-state transition: Both two regions behave as two-state folders but the region of larger $\alpha_i$ folds first and the rest follows. (type V) The three-state transition: At the intermediate state, the region of larger $\alpha_i$ is structurally formed but the rest is unfolded.

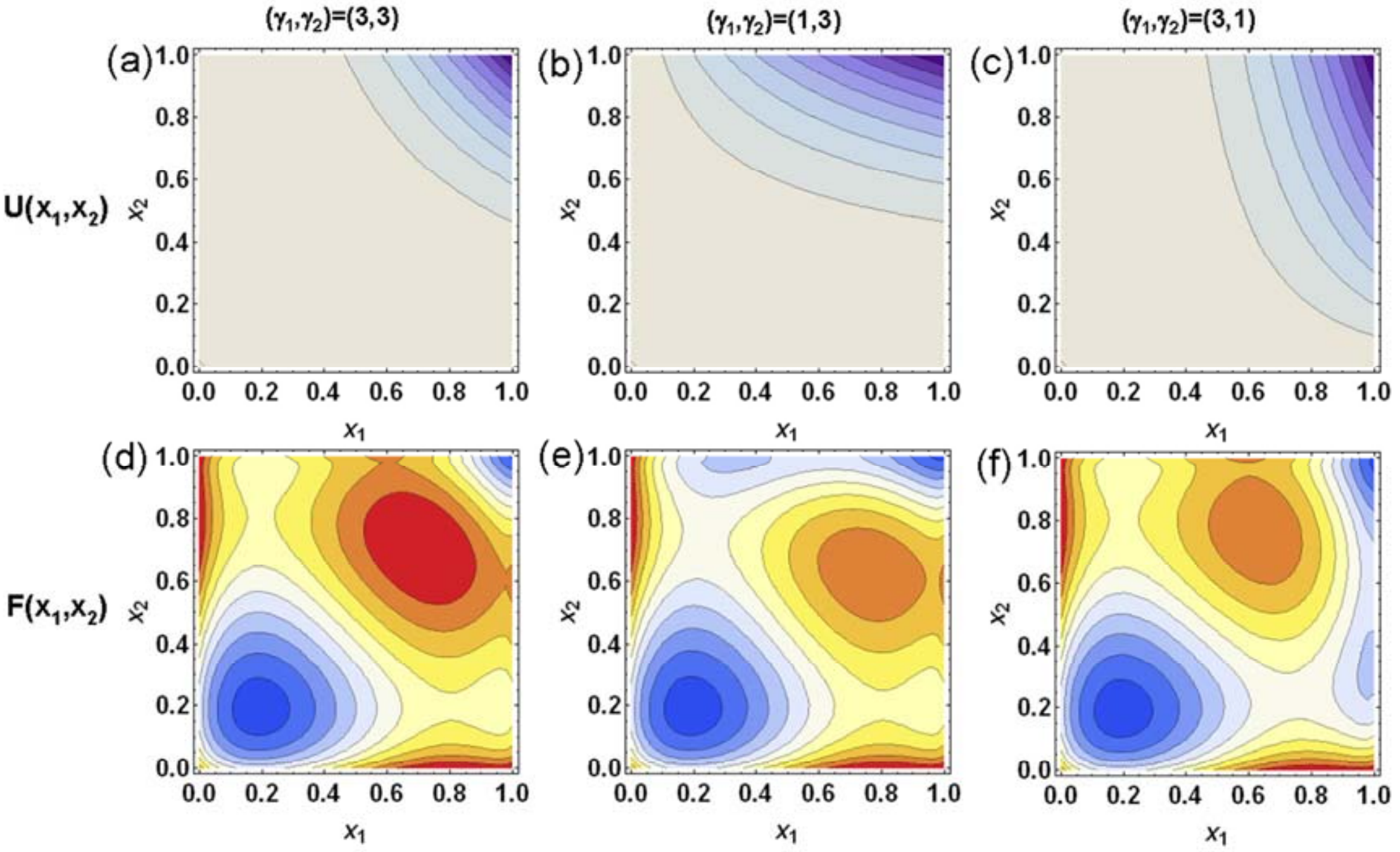

**FIGURE 8** Folding process of case iv classified by the two-dimensional representation of free energy surface. $\alpha_1 = \alpha_2 = 3.5$, $n_{12}/n_{tot} = 0.15$, and $n_1 = n_2$. Shown are $U(x_1,x_2)$ with $(\gamma_1,\gamma_2) = (3,3)$ (a), $(\gamma_1,\gamma_2) = (1,3)$ (b), and $(\gamma_1,\gamma_2) = (3,1)$ (c). $F(x_1,x_2)$ with $(\gamma_1,\gamma_2) = (3,3)$ (d), $(\gamma_1,\gamma_2) = (1,3)$ (e), and $(\gamma_1,\gamma_2) = (3,1)$ (f). In (a-c) contour lines are drawn in every $10k_BT$, and in (d-f) contour lines are drawn in every $2k_BT$.

**Case iv**

This is the case of $\gamma_1 \neq \gamma_2$. In this case $U(x_1,x_2)$ has an asymmetrical functional shape, which changes the statistical weight of two pathways. When $\gamma_i > \gamma_j$, as shown in Fig.8, $U(x_1,x_2)$ enhances the statistical weight of the path through the state in which the region $i$ is more structured than the region $j$. Such asymmetry in the functional form of $U(x_1, x_2)$ may come from the asymmetrical distribution of residues forming native pairs across the region boundary, but it can also arise from the chain connectivity (10): Even when the spatial distribution of residues involved in native pairs is symmetrical, the correlation among interactions should become asymmetrical when the chain asymmetrically connects those residues.

## V. EXAMPLE PROTEINS

### Structure-based Hamiltonian model of folding

In this section we depart from the phenomenological models and deal with a more microscopic structure-based model of folding. We apply the method of multi-dimensional representation discussed in Section III to the free energy calculation with this model.

The free energy surface is calculated by the model introduced by Wako and Saito (51, 52), which was later analyzed extensively by Muñoz and Eaton (53, 54). The model describes the protein conformation with coarse-grained variables $\{m_i\}$. $m_i = 1$ when two dihedral angles of the backbone at the $i$th residue are within some narrow range around values in the native state conformation, and $m_i = 0$ otherwise. We adopt the Hamiltonian:

$$H = -\sum_{i<j} \varepsilon_{i,j} \Delta_{i,j} m_{ij} , \qquad (30)$$

where $m_{ij} \equiv \prod_{k=i}^{j} m_k = m_i m_{i+1} \cdots m_{j-1} m_j$ and $\Delta_{i,j} = 1$ when $i$ and $j$th residues are native pairs, and $\Delta_{i,j} = 0$ otherwise. We define that the pair $i$ and $j$ is a native pair when a heavy atom other than hydrogen in the $i$th residue and a heavy atom in the $j$th residue with $j > i+2$ are closer than 4Å in the native conformation. When the segment of the backbone from $i$ to $j$ takes the native configuration as $m_{ij} = 1$, then the native pair $i$ and $j$ should have a large chance to come close to each other to gain energy of $\varepsilon_{i,j} > 0$. The partition function is calculated as

$$\mathrm{Z} = \mathrm{Tr} \prod_{i=1}^{N} \nu_i^{1-m_i} \, e^{-\beta H}, \qquad (31)$$

where Tr is a sum over $2^N$ possible values of $\{m_i\}$ and $\nu_i$ is the number of nonnative configurations the $i$th residue can take. For simplicity, $\varepsilon_{i,j}$ and $\nu_i$ are assumed to be independent of the residue position: $\varepsilon_{i,j} = \varepsilon$ and $\nu_i = \nu$. Then, the relevant parameters are $\varepsilon/k_B T$ and $\ln \nu$, where $\varepsilon/k_B T$ should take a smaller value when temperature is raised or denaturant is added. $\ln\nu$ is chosen to be 1.5 ($\nu \approx 4.5$) unless otherwise mentioned, but the main results of this section do not depend on the precise value of $\nu$. This Hamiltonian has been applied to folding of many proteins (7, 10, 37. 55-60), to protein mechanical unfolding (61, 62) and to conformational changes in protein functioning (63). Since the model does not include nonnative interactions, this model is a Go-like model which assures the stability of the native conformation for large enough $\varepsilon/k_BT$. With this model, the constrained partition function

$$Z(\mathbf{x}) = \nu^{N(1-\mathbf{x})}\mathrm{Tr}_{\mathbf{x}}\exp(-\beta H), \quad (32)$$

and the constrained average $\langle\cdots\rangle_{\mathbf{x}}$ are calculated without introducing a further approximation (55, 56) (Details are explained in APPENDIX B), so that $F_e^{(i)}(x_i)$, $F_0(\mathbf{x})$, $U(\mathbf{x})$, and $F(\mathbf{x})$ can be exactly calculated from Hamiltonian without introducing the phenomenological assumptions used in previous sections. Here, $x_i = M_i/N_i$ with $M_i = \sum_{k\in i\text{th region}} m_k$, where $\sum_{k\in i\text{th region}}$ denotes the sum of residues within the $i$th region. Eq.32 implies that $S_c^{(i)}(x_i)$ is given by Eq.18.

Before explaining each example protein, we make two remarks on the generic features of the calculated results. First, though $E(\mathbf{x}) = \langle H\rangle_{\mathbf{x}}$ is often a rugged function of $\mathbf{x}$ with many discontinuous changes of its gradient, $F_e(\mathbf{x}) = E(\mathbf{x}) - TS_e(\mathbf{x})$ is a smooth function of $\mathbf{x}$, which implies that the self-consistent compensation of $E(\mathbf{x})$ and $-TS_e(\mathbf{x}) = k_BT\ln\langle\exp[\beta(H-E(\mathbf{x}))]\rangle_{\mathbf{x}}$ is important to determine the free energy surface. When the 1st or 2nd order approximation on the number of contiguous residues taking the native configuration is used, this self-consistent cancelation in the exact calculation is lost and $F(\mathbf{x})$ becomes a rugged function of $\mathbf{x}$. Though this ruggedness was once regarded as an evidence for the incompleteness of the model of Eq.30 (64), it is not the case; Such ruggedness shows the incompleteness of the 1st or 2nd order approximation, and is completely removed from the results of exact calculation. Second, $-TS_e(\mathbf{x})$ is usually large, sometimes amounts to several tens of $k_BT$, and the height and location of transition state ensemble are largely determined by the functional form of $-TS_e(\mathbf{x})$. These two points validate assumptions used in phenomenological models of the preceding sections on the importance of functional forms of $F_e(\mathbf{x})$ and $-TS_e(\mathbf{x})$ or $F_e^{(i)}(x_i)$ and $-TS_{e0}^{(i)}(x_i)$.

Since application of the multi-dimensional representation to multi-domain proteins has been discussed in details in Ref.10, we here mainly focus on the compact single-domain proteins as exemplified below.

**Ribosomal protein S6**

S6 is a 101 residue α+β protein (Protein Data Bank (PDB) code: 1ris), whose structure is shown in Fig.10a. S6 exhibits a classical V-shaped chevron plot for the dependence of observed folding and unfolding rates on denaturant concentration, which implies that S6 is a typical two-state folder (15-18). Both the one-dimensional representation in Fig.9 and the two-dimensional representation in Fig.10 confirm this interpretation: As shown in Fig.9a and Fig.10b, the free energy surface has two minima corresponding to the unfolded and folded states. In one-dimensional representation, we can find that the unfolded state is represented by a large basin of $x < 0.4$. $E(x)$, $-TS_e(x)$, and $F_e(x)$ are flat for $x < 0.2$ as shown in Figs.9b, 9c and 9d, and the sharp increase in $-TS_e(x)$ for $x \approx 0.6$ shown in Fig.9d is the physical origin of a free energy barrier separating the unfolded and folded states in Fig.9a.

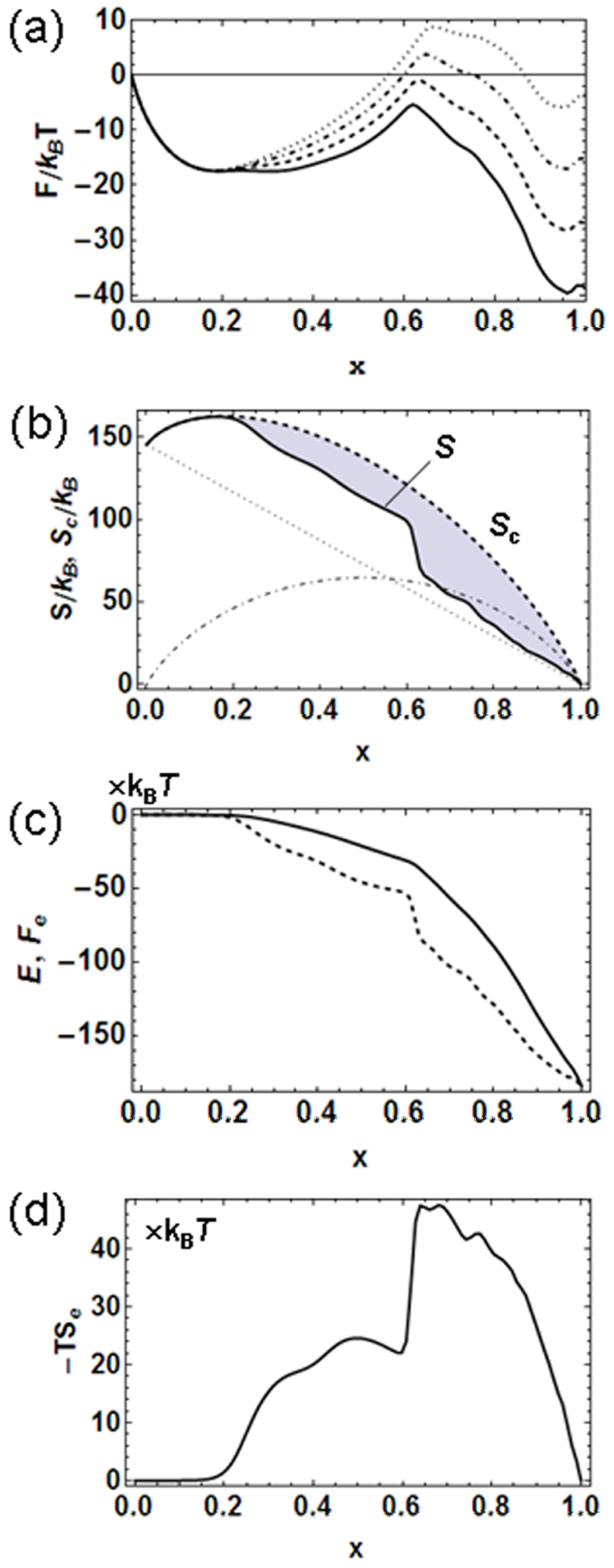

**FIGURE 9**

One-dimensional representation of free energy of ribosomal protein S6. (a) One-dimensional free energy surfaces at $\varepsilon/k_BT$=0.94 (real line), 0.88 (dashed line), 0.82 (dot-dashed line), and 0.76 (dotted line). (b) Total entropy $S(x)$ (real line), conformational entropy $S_c(x)$ (dashed line), $N(k_B \ln \nu)(1-x)$ (dotted line), and $k_B \ln \binom{N}{Nx}$ (dot-dashed line), Difference between $S(x)$ and $S_c(x)$ is $S_e(x)$ (gray area). (c) $F_e(x)$ (real line) and $E(x)$ (dashed line), and (d) $-TS_e(x)$. $\varepsilon/k_BT = 0.94$ for (b)-(d).

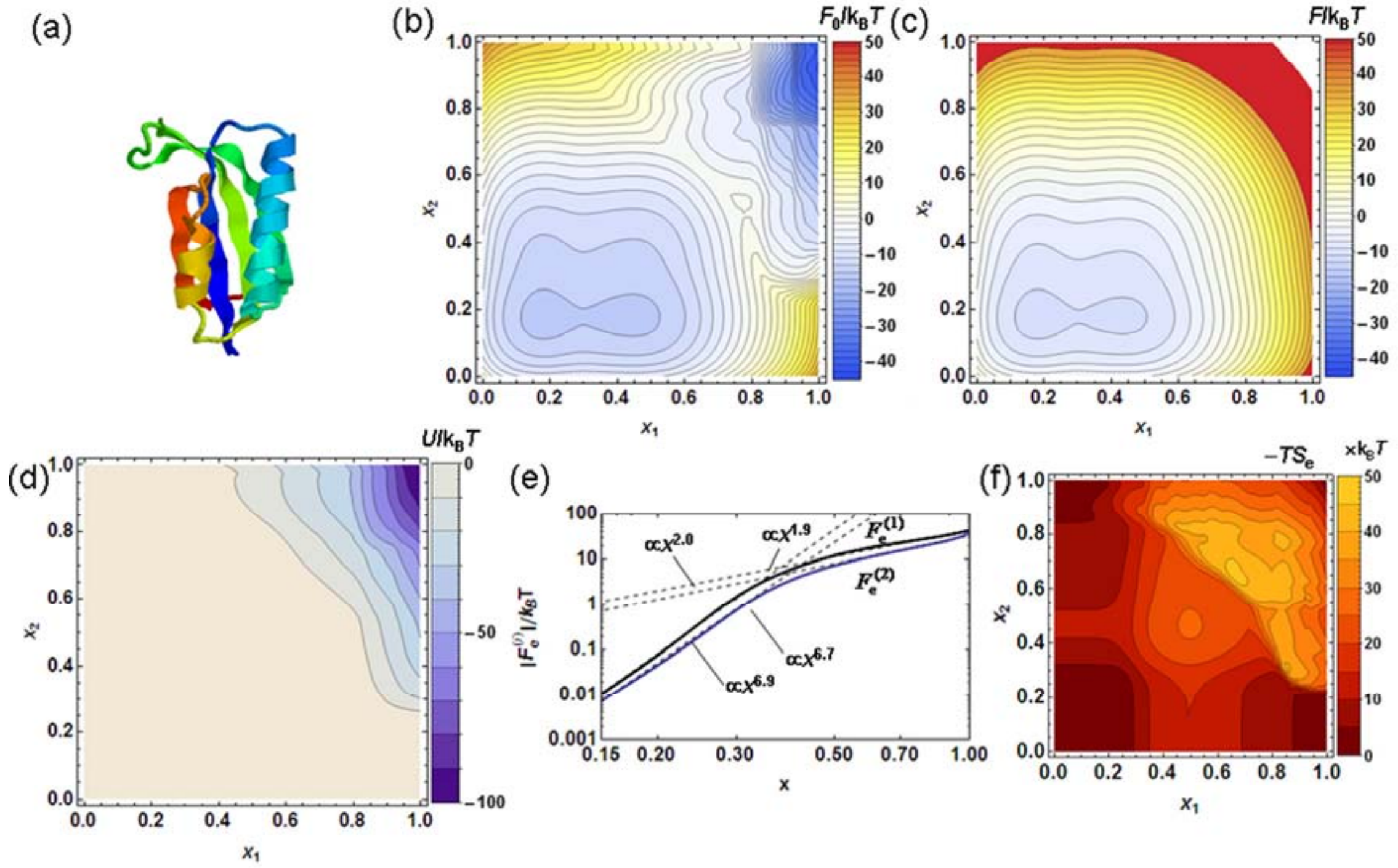

**FIGURE 10** Two-dimensional representation of free energy of ribosomal protein S6. $\varepsilon/k_BT = 0.94$. (a) The native structure of ribosomal protein S6 (PDB ID: 1ris). (b) $F(x_1,x_2)/k_BT$, (c) $F_0(x_1,x_2)/k_BT$, (d) $U(x_1,x_2)/k_BT$, (e) log-log plot of $|F_e^{(1)}(x_1)|/k_BT$ (black line) and $|F_e^{(2)}(x_2)|/k_BT$ (blue line), and (f) $-TS_e(x_1,x_2)$.

Two-dimensional representation is constructed by disregarding the four residues whose structure is experimentally undetermined and by dividing the protein chain with $N_1$ = 49 for the N-terminal region and $N_2$ = 48 for the C-terminal region. Native pairs are distributed to be $(n_1 - n_2)/n_{\mathrm{tot}} = 0.0306$ and $n_{12}/n_{\mathrm{tot}} = 0.56$. The large value of $n_{12}/n_{\mathrm{tot}}$ implies that the each region does not change its structure independently but the two regions behave cooperatively. As shown in Fig.10e $F_e^{(i)}(x_i)$ can be fitted as $F_e^{(1)}(x_1) \propto -1/\{(x_1/x_0)^{-6.9} + (x_1/x_0)^{-1.9}\}$ and $F_e^{(2)}(x_2) \propto -1/\{(x_2/x_0)^{-6.7} + (x_2/x_0)^{-2.0}\}$ with $x_0 \approx 0.4$. Since $x_i < x_0$ is the state of a flat unfolded basin, we can focus on the case of $x_i > x_0$, where $F_e^{(i)}(x_i)$ can be represented by $F_e^{(i)}(x_i) \propto -x_i^{\alpha_i}$ with $\alpha_1 = 1.9$ and $\alpha_2 = 2.0$. Cooperativity within each region is weak due to these small $\alpha_i$, and the folded state is strongly stabilized by $U(x_1,x_2)$ as expected from the large $n_{12}/n_{\text{tot}}$. As $(n_1-n_2)/n_{\text{tot}}$ has some nonzero value and $\alpha_1 \approx \alpha_2 \approx 2.0$ and $n_{12}/n_{\text{tot}} > 0.5$, S6 resembles to **type IIa of case ii** in the phenomenological model discussed in Section IV. The saddle point locates at $x_1 \approx x_2$, which is consistent with the observed large Φ-value at the middle of the chain (16,17).

When the structure of the mutant L30A (PDB code: 1lou) is used instead of the wild type S6, The folded state is destabilized by $6k_BT$ (≈ 6.4ε) increase in free energy when the same parameter set of $\varepsilon/k_BT$ and $\ln\nu$ are used. The two-dimensional representation in Fig.11 shows that a shallow minimum at a depth of about $2k_BT$ appears near the saddle, which is consistent with the observation of the curved chevron plot at the high denaturant concentration side (16,19). This additional small minimum is undetectable in the one-dimensional representation because the minimum is masked by many other states projected together onto a single coordinate.

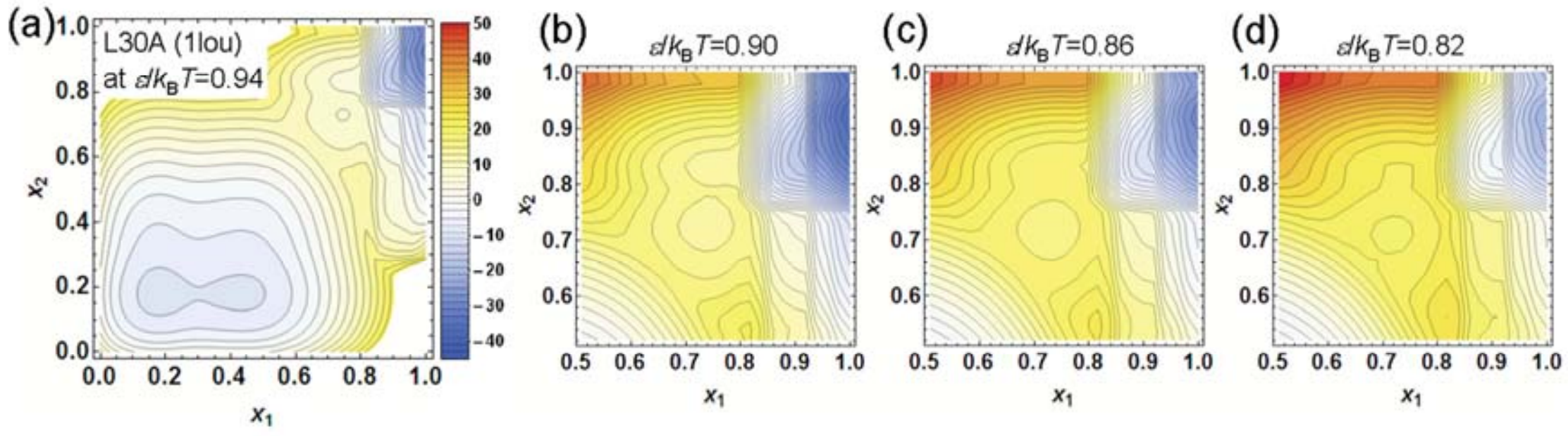

**FIGURE 11** Two-dimensional representation of free energy of the L30A mutant of ribosomal protein S6. (a) $F(x_1,x_2)/k_BT$ of the L30A mutant at the folding condition with $\varepsilon/k_BT = 0.94$. Shown are $F(x_1,x_2)/k_BT$ of the L30A mutant at $\varepsilon/k_BT = 0.90$ (b), 0.86 (c), and 0.82(d). In (b-d) contour lines are drawn in every $k_BT$.

**Src-SH3 domain**

The src-SH3 domain, whose structure is shown in Fig.12a, is a 57-residue domain (PDB code:1fmk) exhibiting the two-state folding transition (20,21). By dividing the chain with $N_1$ = 29 for the N-terminal region and $N_2$ = 28 for the C-terminal region, the two-dimensional representation of free energy surface is calculated as shown in Fig.12. Native pairs are distributed to be $(n_1-n_2)/n_{\text{tot}} = 0.077$ and $n_{12}/n_{\text{tot}} = 0.42$. $F_e^{(i)}(x_i)$ can be fitted as $F_e^{(1)}(x) \approx F_e^{(2)}(x) \propto -1/\{(x/x_0)^{-6.6} + (x/x_0)^{-2.45}\}$ with $x_0 \approx 0.6$, and $F_{\mathrm{e}}^{(i)}(x_i) \propto -x_i^{\alpha_i}$ for large $x_i$ with $\alpha_1 \approx \alpha_2 \approx 2.45$. Since $\alpha_i$ is rather large, each region itself tends to fold in a two-state manner, but the large $n_{12}/n_{\mathrm{tot}}$ inhibits independent folding of two regions: The whole protein exhibits the two-state transition without being trapped at the intermediate state. Src-SH3 resembles to **type IIb of case ii** in the phenomenological model of Section IV. From Fig.12b, we can see that there are two folding pathways, one passing the state of large $x_2$ and small $x_1$, and the other passing the state of large $x_1$ and small $x_2$: Along the former path, which we call $\mathrm{path_{CN}}$, the C-terminal half folds first and catalyzes folding of the N-terminal half, and along the latter path, which we call $\mathrm{path_{NC}}$, the N-terminal half folds first and catalyzes folding of the C-terminal half. Since $n_1 > n_2$, the N-terminal region is more stable than the C-terminal region, which may stabilize $\mathrm{path_{NC}}$, but the large negative values of asymmetric function $U(x_1, x_2)$ in Fig.12c overcome this effect and stabilize $\mathrm{path_{CN}}$ more than $\mathrm{path_{NC}}$. Existence of multiple pathways is consistent with the data derived from simulations with the off-lattice Go model, which suggests the heterogeneity of the transition state ensemble (22).

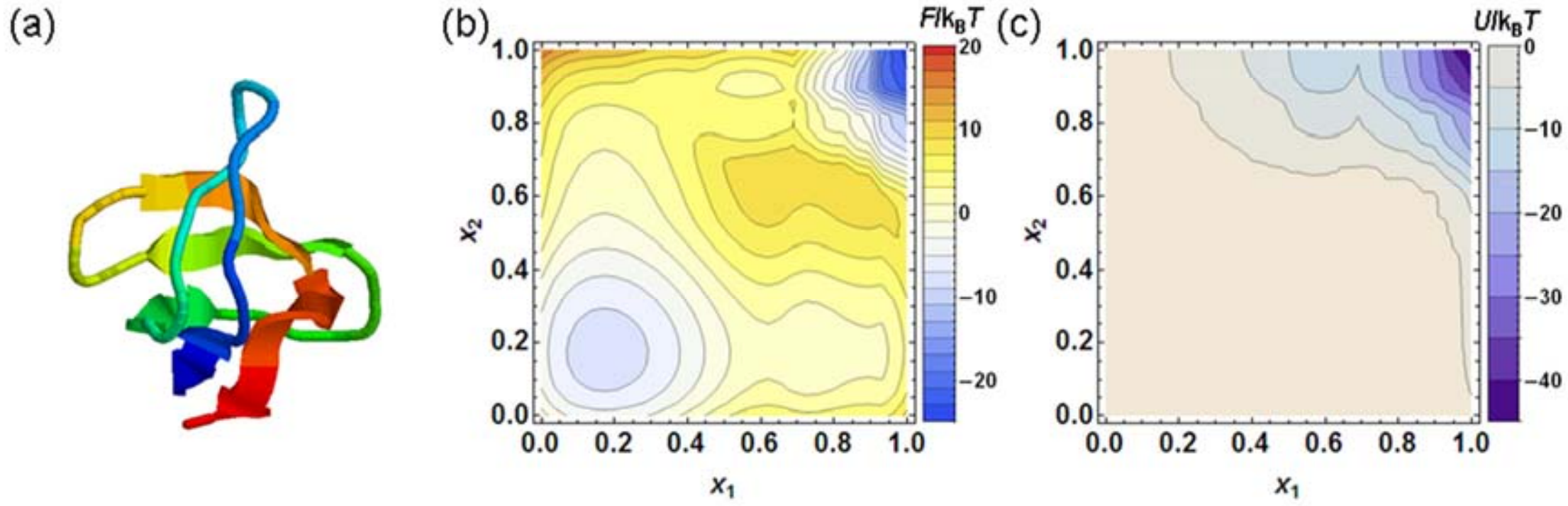

**FIGURE 12** Two-dimensional representation of free energy of src-SH3 domain. $\varepsilon/k_{\mathrm{B}}T$ = 1.05. (a) The native structure of src-SH3 domain (PDB ID: 1fmk). (b) $F(x_1,x_2)/k_{\mathrm{B}}T$, and (c) $U(x_1,x_2)/k_{\mathrm{B}}T$.

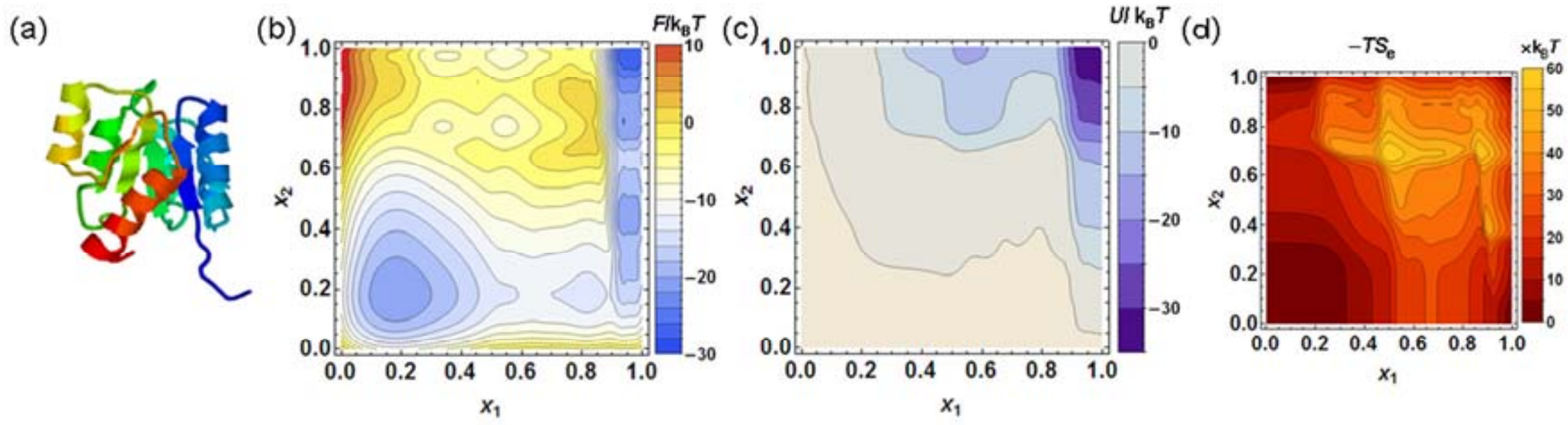

**FIGURE 13** Two-dimensional representation of free energy of CheY. $\varepsilon/k_BT = 0.78$. (a)The native structure of CheY (PDB ID: 3chy). (b) $F(x_1,x_2)/k_BT$, (c) $U(x_1,x_2)/k_BT$, and (d) $-TS_e(x_1,x_2)$.

**CheY**

CheY is a 129-residue α/β parallel protein (PDB code: 3chy), whose structure is shown in Fig.13a, exhibiting a kinetic intermediate in the folding process (23). H/D exchange and NMR data showed that in this intermediate the first α helix and the first two β sheets in the N-terminal half is structured but the C-terminal half remains unstructured though the native conformation shows CheY is a single-domain protein (23-25). The two-dimensional representation is constructed by disregarding one residue whose configuration is undetermined in the native state and by dividing the chain with $N_1 = 64$ for the N-terminal region and $N_2 = 64$ for the C-terminal region. $(n_1 - n_2)/n_{\mathrm{tot}} = 0.057$ and $n_{12}/n_{\mathrm{tot}} = 0.16$, and $F_{\mathrm{e}}^{(i)}(x_i)$ can be fitted as $F_{\mathrm{e}}^{(i)}(x_i) \propto -x_i^{\alpha_i}$ for large $x_i$ with $\alpha_1 \approx \alpha_2 \approx 3.0$. The free energy surface of Fig.13b has a distinct minimum at $x_1 \approx 1.0$ and $x_2 \approx 0.4$, which is consistent with the observed feature of the kinetic intermediate. $U(x_1,x_2)$ in Fig.13c, which has a large negative value and gives rise to a deep minimum in $F(x_1,x_2)$ at the folded state, also exhibits the low free energy valley along $x_1 \approx 1.0$. This asymmetric functional form of $U(x_1,x_2)$ is the origin of the predominance of the pathway via the structured N-terminal half and the unstructured C-terminal half. Along this pathway, a minimum representing the intermediate appears due to the superposition of a local minimum of $-TS_e(x_1, x_2)$ on the flat low free energy valley of $U(x_1,x_2)$, where the functional form of $-TS_e(x_1, x_2)$ is shown in Fig.13d. Thus, CheY resembles to **type IIIa of case ii**, but its mechanism of how the predominant pathway is selected is largely owing to the asymmetric functional form of $U(x_1,x_2)$, which was not classified in cases discussed in Section IV. When we divide the chain with $N_1 = 85$ for the N-terminal region and $N_2 = 43$ for the C-terminal region, however, $F_0(x_1,x_2)$ has an asymmetric functional form with a minimum at $x_1 \approx 0.98$ and $x_2 \approx 0.18$ corresponding to the intermediate. In the vicinity of the intermediate, $U(x_1, x_2)$ does not have an influence on the functional form of $F(x_1,x_2)$. In the case of selecting $N_1$=85 and $N_2$=43, in which $(n_1 - n_2)/n_{\rm tot} = 0.418$ and $n_{12}/n_{\rm tot} = 0.139$, CheY is more precisely regarded to be type IIIa of case ii.

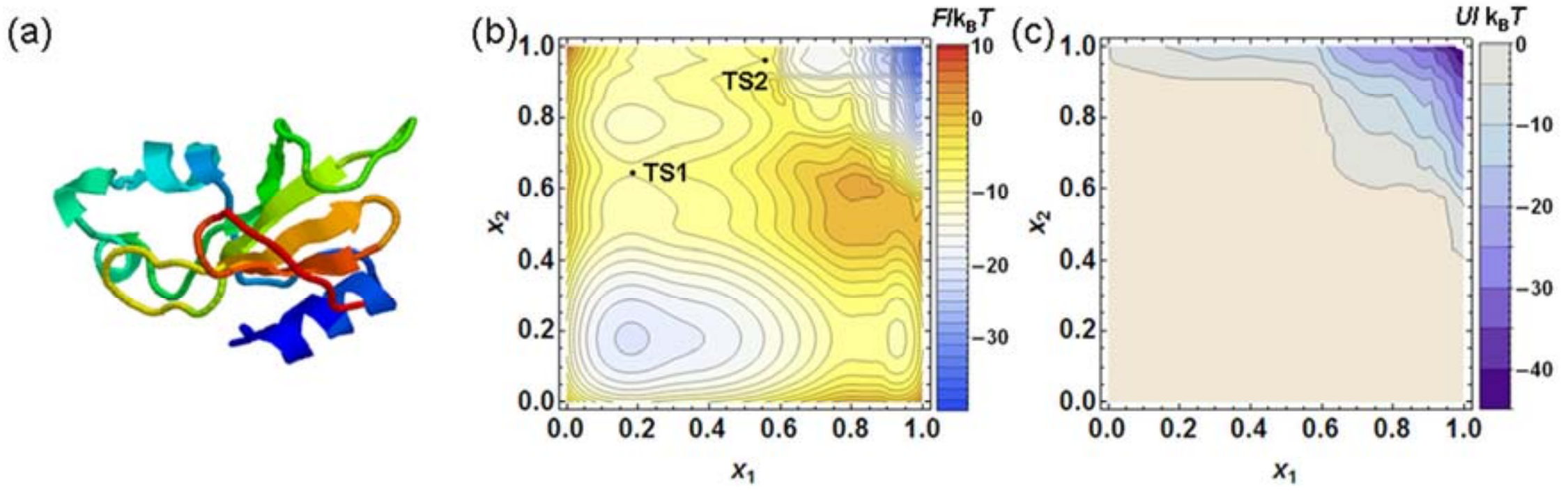

**FIGURE 14** Two-dimensional representations of free energy of Barnase. $\varepsilon/k_B T$ = 0.82. (a) The native structure of Barnase (PDB ID: 1a2p). (b) $F(x_1,x_2)/k_B T$, and (c) $U(x_1,x_2)/k_B T$

**Barnase**

Barnase is a 110 residue α+β protein (PDB code: 1a2p), which exhibits a high-energy intermediate in the folding process (26), whose structure is shown in Fig.14a. Both Φ-values for the transition from the unfolded state to the intermediate and Φ-values for the transition from the intermediate to the folded state have been observed experimentally, which has characterized the structure features of the transition state ensemble between the unfolded state and the intermediate (TS1) and that between the intermediate and the folded state (TS2) (27, 28). The two-dimensional representation is constructed by disregarding two structurally undetermined residues and dividing the chain with $N_1$ = 54 for the N-terminal region and $N_2$ = 54 for the C-terminal region. The free energy surface of Fig.14b has a distinct minimum at $x_1 \approx 0.2$ and $x_2 \approx 0.8$, which corresponds to the kinetic intermediate. $(n_1 - n_2)/n_{\rm tot} = 0.0086$ and $n_{12}/n_{\rm tot} = 0.24$. $F_{\rm e}^{(i)}(x_i)$ can be fitted as $F_{\rm e}^{(i)}(x_i) \propto -x_i^{\alpha_i}$ with $\alpha_1 \approx 3.0$ for $x_1 > 0.4$ and with $\alpha_2 \approx 4.0$ for $x_2 > 0.6$ and $\alpha_2 \approx 9.5$ for $0.5 < x_2 < 0.6$. The larger value of $\alpha_2$ than $\alpha_1$ implies that the C-terminal half folds more cooperatively with larger $-TS_{\rm e}$ than the N-terminal half, which is the reason why the minimum at $x_1 \approx 0.2$ and $x_2 \approx 0.8$ is more distinct than the additional minimum at $x_1 \approx 0.9$ and $x_2 \approx 0.2$. Considering that the difference between $n_1$ and $n_2$ is much smaller than the other example proteins tested in this paper, we can regard that barnase is **type V' of case iii**. As shown in Fig.14c, $U(x_1, x_2)$ is asymmetric and lowers the free energy of TS2.

With the two-dimensional representation, one can calculate Φ-values by sampling conformations at the saddle of the two-dimensional free energy surface. In Fig.15, Φ-values thus calculated for TS1 and TS2 are compared with the experimental data (27,28). Here, Φ-value was defined as follows: When the strength of interactions which involve the $i$th residue is changed from $\varepsilon$ to $\varepsilon + \Delta\varepsilon$, then energy is modulated as $\Delta H(i) = \Delta\varepsilon \sum_j \Delta_{i,j} m_{ij}$. Φ-value of the $i$th residue is obtained by

$$\Phi_i = \frac{-k_B T \ln\langle \exp[-\beta \Delta H(i)] \rangle_{TS}}{\Delta\varepsilon \sum_j \Delta_{i,j}}, \qquad (33)$$

where $\langle \cdots \rangle_{TS}$ is the average taking under the constraint of the transition state (TS1 or TS2).

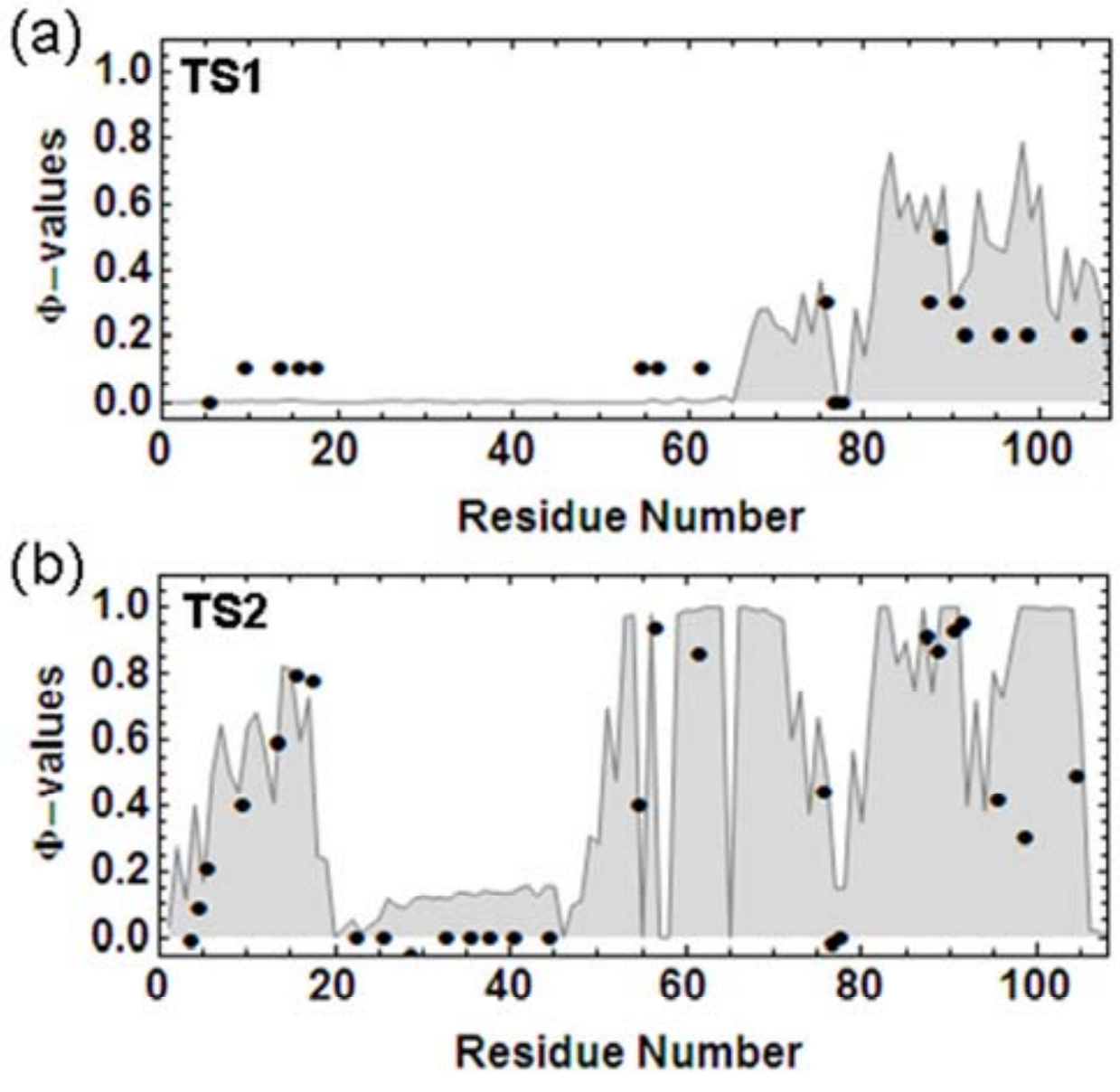

**FIGURE 15** Φ-values of Barnase calculated using $\Delta\varepsilon = -0.01 k_B T$ in Eq.33 are compared with the observed data taken from Refs.27 and 28. (a) Φ-values at the transition state between the unfolded and intermediate states (TS1). Data connected by a line are Φ-values calculated at $(x_1, x_2) = (0.185, 0.648)$, and dots are observed values. (b) Φ-values at the transition state between the intermediate and folded states (TS2). Data connected by a line are Φ-values calculated at $(x_1, x_2) = (0.556, 0.945)$, and dots are observed values.

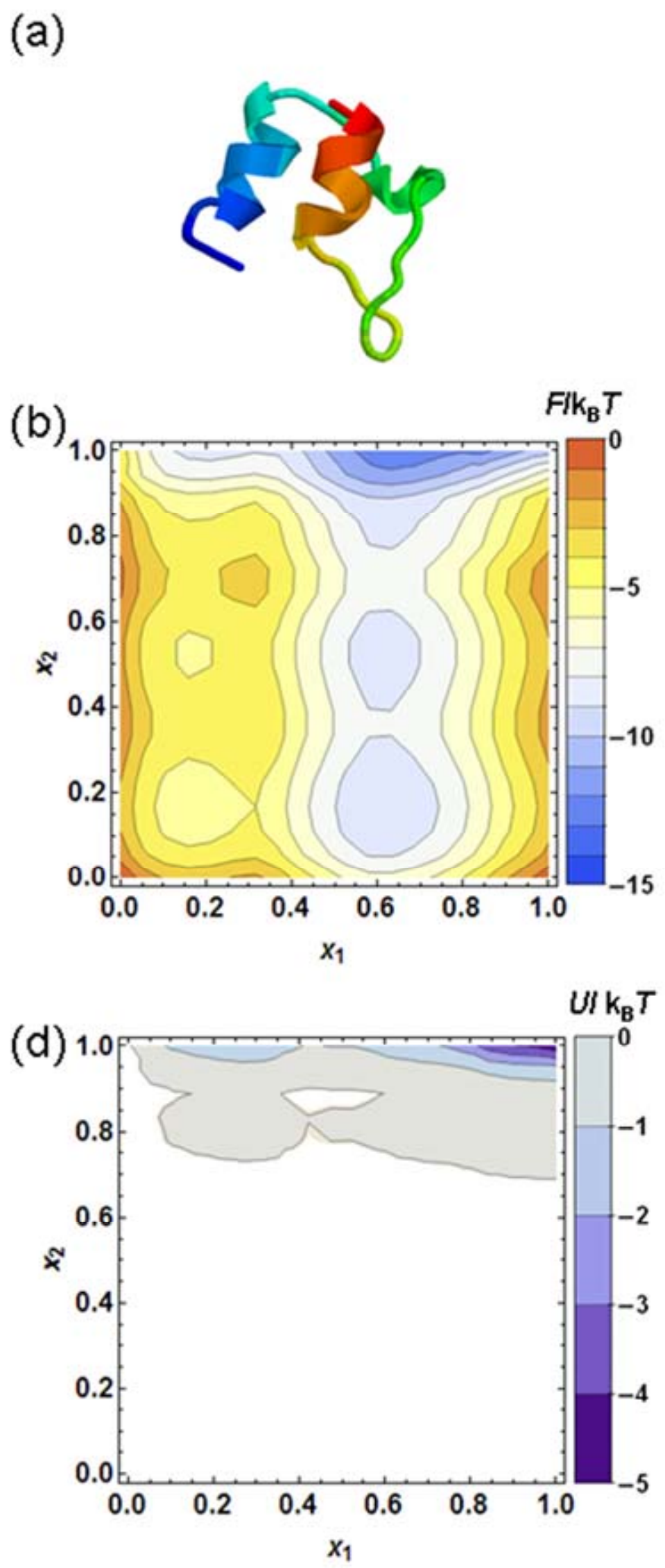

**FIGURE 16** Two-dimensional representation of free energy of BBL domain. $\varepsilon/k_BT$ = 1.60. (a) The native structure of BBL domain (PDB ID: 1bbl). (b) $F(x_1,x_2)/k_BT$, and (c) $U(x_1,x_2)/k_BT$.

**BBL domain**

BBL is a domain consisting of about 40 residues excised from a multi-enzyme complex. Variants of BBL structure have been studied by different groups (29-32) and we here use the structure determined by NMR (PDB code: 1bbl), whose structure is shown in Fig.16a. Though the sequence is composed of 50 residues in this PDB record, 10 residues at the N-terminus and 3 residues at the C terminus are unstructured and only 37 residues are structured. The two-dimensional representation is constructed by dividing the structured part of the chain as $N_1$ = 19 and $N_2$ = 18. As shown in Fig.16b, there are minima in the free energy surface, corresponding to the unfolded state ($x_1 \approx 0.2$, $x_2 \approx 0.2$), the intermediate state ($x_1 \approx 0.6$, $x_2 \approx 0.2 - 0.5$), and the folded state ($x_1 \approx 0.7$, $x_2 \approx 1.0$). Though there is a free energy barrier between the intermediate state and the folded state, the barrier height is as small as $1k_\mathrm{B}T$. The barrier height between the intermediate and the unfolded state is even smaller. This weak two- or three-state feature can be confirmed also in the one-dimensional representation of Fig.17a for ln$\nu$ = 1.5. One-dimensional $F_\mathrm{e}(x)$ for ln$\nu$ = 1.5 can be fitted as $F_\mathrm{e}(x) \propto -x^\alpha$ with $\alpha = 2.4$ for $x \geq 0.5$, showing that BBL domain resides at around the boundary between two-state folding and downhill folding at $\alpha \approx \alpha_\mathrm{c}$ (Fig.1a). As shown in Fig.17b the one-dimensional free energy surface for ln$\nu$ = 1.0 indicates a barrier less downhill folding. $U(x_1, x_2)$ in Fig.16c, which is asymmetric and lowers the free energy at $x_2 \approx 1$, shows that inter-region interactions only catalyze folding of N-terminal half with the folded C-terminal half. $(n_1 - n_2)/n_\mathrm{tot} = -0.025$ and $n_{12}/n_\mathrm{tot} = 0.075$. $F_\mathrm{e}^{(i)}(x_i) \propto -x_i^{\alpha_i}$ with $\alpha_1 \approx 1.9$ and $\alpha_2 \approx 2.7$, so that BBL is **type III'b of case iii**, but is close to the border of type I. If we regard $\alpha_1 \approx \alpha_2$, BBL can be also mapped to **type II of case ii**. In this classification, BBL locates close to the border of type I of case ii, so that small increase of $n_{12}/n_\mathrm{tot}$ should make BBL a more distinct two-state folder, and small decrease of $n_{12}/n_\mathrm{tot}$ should make BBL a downhill folder. Such sensitivity to $n_{12}/n_\mathrm{tot}$ is consistent with the simulated sensitivity of folding features to the minor modification of structure (33) and can explain the seeming contradiction among observed results (34-36).

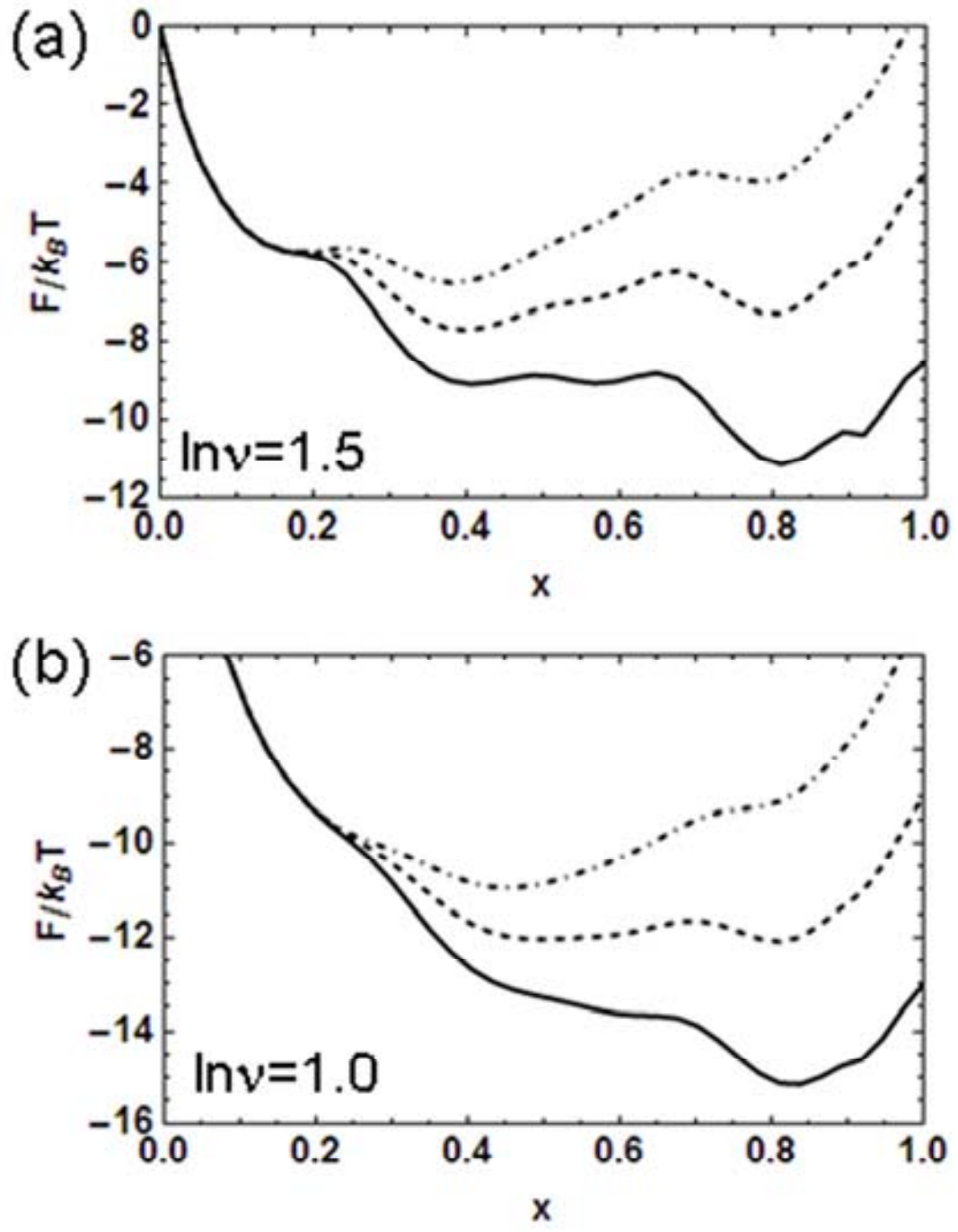

**FIGURE 17** One-dimensional representation of free energy of BBL domain. (a) $F(x)$ with $\ln\nu = 1.5$ at $\varepsilon/k_\mathrm{B}T$ = 1.60 (real line), 1.48 (dashed line), and 1.37 (dot-dashed line). (b) $F(x)$ calculated with $\ln\nu = 1.0$ at $\varepsilon/k_\mathrm{B}T$ = 1.60 (real line), 1.48 (dashed line), and 1.37 (dot-dashed line). Height of the free energy barrier separating the unfolded and folded states becomes less than $k_\mathrm{B}T$.

## VI. EXTENSION TO HIGHER DIMENSION

It is straightforward to develop the three or higher dimensional representation of free energy surface by using the method described in Section III. A way to visualize the results is to draw suitable two-dimensional cross sections of the higher dimensional surface. For the $L$-dimensional surface with $\mathbf{x} = (x_1, x_2, \cdots, x_L)$, a cross section is obtained by using the two-dimensional coordinate ($x_i$,$x_j$) and keeping the other coordinates constant as $x_k = x_{k0}$ for $k \neq i$ or $j$. In this section, examples of two-dimensional cross sections of the three-dimensional free energy surface are discussed by calculating free energies of several proteins with the same structure-based Hamiltonian model as used in the last section.

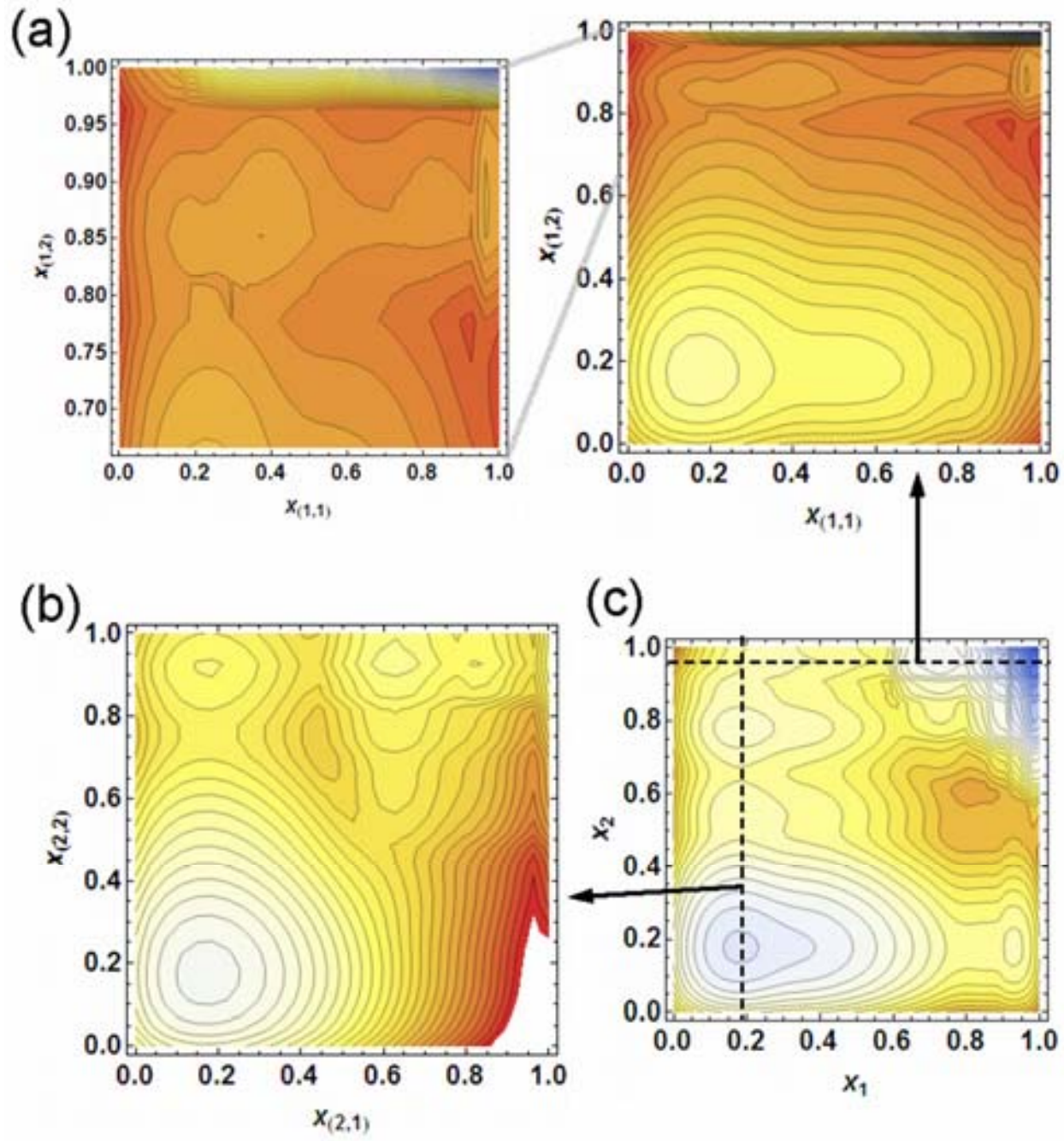

**FIGURE 18** Two-dimensional cross-sections of three-dimensional representations of free energy of barnase. (a) The cross-section at ($x_{(1,1)}$, $x_{(1,2)}$, $x_2$ = 0.94), and (b) the cross-section at ($x_1$ = 0.18, $x_{(2,1)}$, $x_{(2,2)}$). In (a) and (b) contour lines are drawn in every $k_BT$. (c) Two-dimensional representation.

### Barnase

Examples of two-dimensional cross sections of the three-dimensional surface of barnase are shown in Fig.18. We can define three regions, for example, by dividing the 2nd region of barnase into region(2,1) and region(2,2), where the 2nd region is the C-terminal half of barnase with $N_2$ = 54 as explained in Section V, region(2,1) is the N-terminal half of the 2nd region, which consists of $N_{(2,1)}$ = 27 residues, and region(2,2) is the rest half of the 2nd region, which consists of $N_{(2,2)}$ = 27 residues. We use the 1st region, which is the N-terminal half of barnase with $N_1$ = 54, as explained in Section V. The three-dimensional coordinate is defined by $\mathbf{x} = (x_1, x_{(2,1)}, x_{(2,2)})$, where $x_1 = M_1/N_1$ as before and $x_{(2,1)} = M_{(2,1)}/N_{(2,1)}$ and $x_{(2,2)} = M_{(2,2)}/N_{(2,2)}$. $M_{(2,1)}$ is the number of residues taking the native-like configuration in region(2,1), and $M_{(2,2)}$ is the number of residues taking the native-like configuration in region(2,2). The most informative intersection is obtained by setting $x_1 = 0.18$, which is a line connecting the unfolded and intermediate states in the $(x_1, x_2)$ representation of Fig.14b or Fig.18c. The free energy surface of the cross section with $(x_1 = 0.18, x_{(2,1)}, x_{(2,2)})$ is shown in Fig.18b. There exist two parallel pathways from the unfolded state at $x_{(2,1)} \approx 0.2$ and $x_{(2,2)} \approx 0.2$ to the intermediate state at $x_{(2,1)} \approx 0.65$-$0.8$ and $x_{(2,2)} \approx 0.9$. One pathway has a kinetic intermediate state, in which region(2,2) is folded and region(2,1) is unstructured. Along another pathway, region(2,1) and region(2,2) fold concurrently. The statistical weight of these two pathways should be similar to each other as the height of the rate-limiting barrier is almost the same along two pathways. In this way, the fine features of heterogeneity of transition states and variety of pathways are more visible with the three-dimensional representation.

Also shown in Fig.18a is another three-dimensional representation obtained by dividing the 1st region into region(1,1) and region(1,2) with $N_{(1,1)} = 27$ and $N_{(1,2)} = 27$. The three-dimensional coordinate is defined by $\mathbf{x} = (x_{(1,1)}, x_{(1,2)}, x_2)$ with $x_{(1,1)} = M_{(1,1)}/N_{(1,1)}$ and $x_{(1,2)} = M_{(1,2)}/N_{(1,2)}$. Its two-dimensional cross section at $x_2 = 0.94$ is shown in Fig.18a, which represents the transition from the intermediate to folded states. In Fig.18a, the saddle at TS2 between the intermediate and folded states of the two-dimensional surface in Fig.14b is expanded to be a broad area having two local minima at $(x_{(1,1)}, x_{(1,2)}) \approx (0.3, 0.85)$ and $(0.8, 0.85)$. It is interesting to see that the basin of the folded state elongates toward the small $x_{(1,1)}$ direction in Fig.18a, implying that region(1,1), which includes the N-terminal helix, fluctuates more largely in the native state than the rest part. Weaker stability of the N-terminal helix in the native state and the fast folding of the N-terminal helix are compatible in the nonlinearly curved free energy landscape, which was not visible in one or two dimensional representation of section V. The flexible N-terminal helix should explain the locating process of barnase *in vivo*: It has been experimentally observed that unfolding of the N-terminal helix initiates importing barnase into a mitochondrion (65), and the intrinsic flexibility of the N-terminal helix shown in the present three-dimensional representation is consistent with this observation.

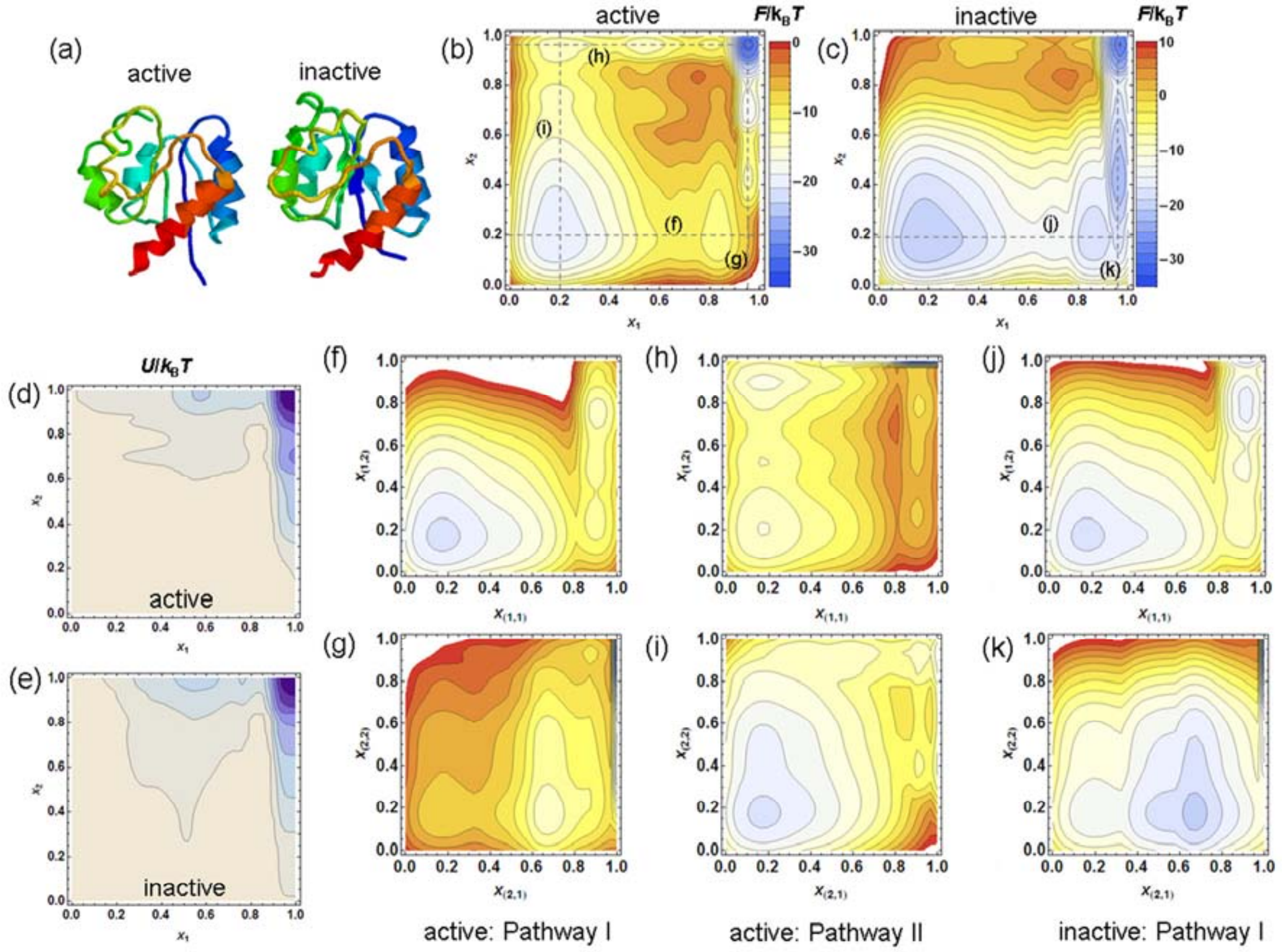

**FIGURE 19** Two-dimensional cross-sections of three-dimensional representations of free energy of active phosphorylated and inactive dephosphorylated NtrC. (a) The native structure of active NtrC (PDB ID: 1dc8) and inactive NtrC (PDB ID: 1dc7). Two-dimensional representations of free energy of the active (b) and the inactive (c) structures. Two-dimensional representations of $U(x_1,x_2)/k_BT$ of the active (d) and the inactive (e) structures. Shown are the cross-section of the three-dimensional free energy surface of the active structure at ($x_{(1,1)}$, $x_{(1,2)}$, $x_2$ = 0.20) (f) and ($x_1$ = 0.95, $x_{(2,1)}$, $x_{(2,2)}$) (g) for Pathway I, and at ($x_{(1,1)}$, $x_{(1,2)}$, $x_2$ = 0.95) (h), and ($x_1$ = 0.20, $x_{(2,1)}$, $x_{(2,2)}$) (i) for Pathway II. The cross-section of the the three-dimensional free energy surface of the inactive structure at ($x_{(1,1)}$, $x_{(1,2)}$, $x_2$ = 0.20) (j) and ($x_1$ = 0.95, $x_{(2,1)}$, $x_{(2,2)}$) (k). In (d-e) contour lines are drawn in every $5k_BT$ and in (f-k) contour lines are drawn in every $2k_BT$. $\varepsilon/k_BT = 0.9$.

## NtrC

Fig.19 shows examples of free energy surface calculated for the receiver domain of the bacterial enhancer-binding protein NtrC (38, 39). The receiver domain NtrC is a single-domain α/β protein with 124 residues and its inactive structure (PDB ID: 1dc7) is switched to the active structure (PDB ID: 1dc8) by phosphorylation (38). Two-dimensional representation of free energy of the active phosphorylated (Fig. 19b) and the inactive dephosphorylated (Fig.19c) NtrC is constructed by dividing the chain with $N_1$ = 62 for the N-terminal region and $N_2$ = 62 for the C-terminal region. In the active NtrC there are two folding pathways as shown in Fig. 19b, each of which passes its specific local minimum: One is the pathway along which the N-terminal half folds first and the C-terminal half follows, which is denoted here by Pathway I, and along the other pathway, the C-terminal half folds first and the N-terminal half follows, which is denoted by Pathway II. In contrast, as shown in Fig.19c, the inactive NtrC has only one dominant folding pathway, Pathway I. This difference is due to the difference in sign of $(n_1 - n_2)/n_{\mathrm{tot}}$ : $(n_1 - n_2)/n_{\mathrm{tot}} = -0.047$ in the active NtrC and $(n_1 - n_2)/n_{\mathrm{tot}} = 0.098$ in the inactive NtrC. In the active NtrC, $n_2 > n_1$ makes Pathway II the most dominant pathway and Pathway I remains the next dominant pathway. In the inactive NtrC, the dominant folding pathway is Pathway I because of $n_1 > n_2$. Both in the active and inactive NtrC, as shown in Fig. 19d and 19e, $U(x_1, x_2)$ has asymmetrical functional forms, which lowers the free energy along Pathway I. This asymmetry of $U(x_1, x_2)$ makes the difference between Pathway I and Pathway II small in the active NtrC, but the asymmetry enlarges the difference between two pathways in the inactive NtrC. Other indices are almost same for the active and inactive NtrC: $n_{12}/n_{\mathrm{tot}}$ =0.16 in the active NtrC and $n_{12}/n_{\mathrm{tot}}$ =0.18 in the inactive NtrC. In both structures, $F_{\mathrm{e}}^{(i)}(x_i)$ can be fitted as

$F_{\mathrm{e}}^{(i)}(x_i) \propto -x_i^{\alpha_i}$ for large $x_i$ with $\alpha_1 \approx \alpha_2 \approx 3.0$.

Two-dimensional cross sections of three-dimensional surfaces of the active NtrC, ($x_{(1,1)}$, $x_{(1,2)}$, $x_2$ = 0.20), ($x_1$ = 0.95, $x_{(2,1)}$, $x_{(2,2)}$), ($x_{(1,1)}$, $x_{(1,2)}$, $x_2$ = 0.95), and ($x_1$ = 0.20, $x_{(2,1)}$, $x_{(2,2)}$), are shown in Figs. 19f, 19g, 19h and 19i, and those of the inactive NtrC, ($x_{(1,1)}$, $x_{(1,2)}$, $x_2$ = 0.20) and ($x_1$ = 0.95, $x_{(2,1)}$, $x_{(2,2)}$), are shown in Figs. 19j and 19k, respectively with $N_{(1,1)}$ = 31, $N_{(1,2)}$ = 31, $N_{(2,1)}$ = 31, and $N_{(2,2)}$ = 31. These free energy surfaces show that the cooperativity in folding is pathway-dependent: The N-terminal half folds as region(1,2) → region(1,1) when the C-terminal half is folded (Fig. 19h), whereas it folds as region(1,1) → region(1,2) when the C-terminal half is unfolded (Fig. 19f or 19j). The C-terminal half folds as region(2,1) → region(2,2) when the N-terminal half is folded (Fig. 19g or 19k), whereas it folds as region(2,2) → region(2,1) when the N-terminal half is unfolded (Fig.19i). Summarizing these observations, we can conclude that along Pathway I, the structure is sequentially constructed from the N-terminal to C-terminal regions as region(1,1) → region(1,2) → region(2,1) → region(2,2), whereas along Pathway II the structure is sequentially constructed from the C-terminal to N-terminal regions as region(2,2) → region(2,1) → region(1,2) → region(1,1).

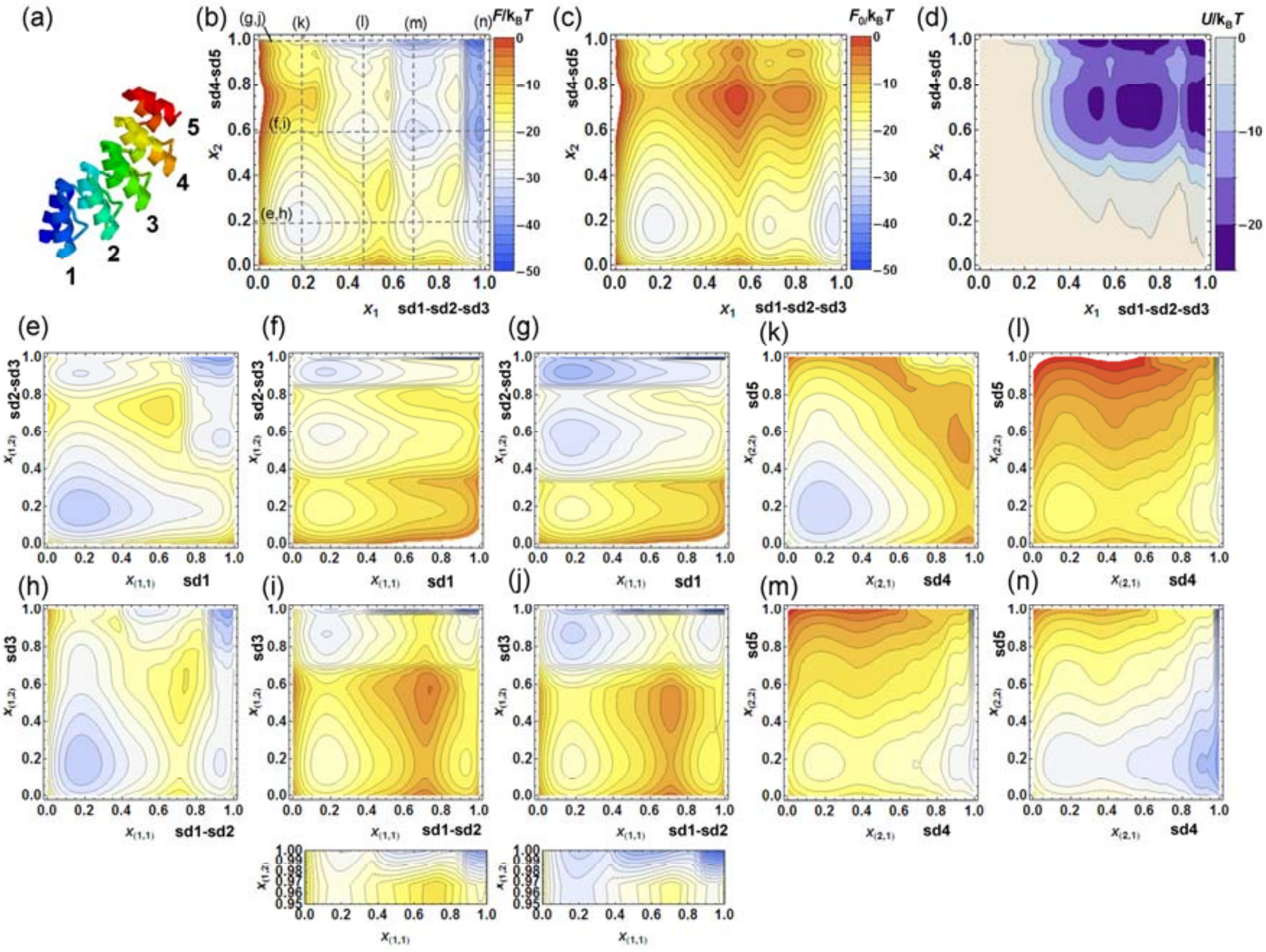

**FIGURE 20** Two-dimensional cross-sections of three-dimensional representations of free energy of an ankyrin repeat protein. (a) The native structure of the repeat protein (PDB ID: 1mj0) which has five sub-domains (sd1, sd2, sd3, sd4, and sd5). Two-dimensional representations of (b) $F(x_1,x_2)/k_BT$, (c) $F_0(x_1,x_2)/k_BT$, and (d) $U(x_1,x_2)/k_BT$. Shown are the cross-section of the three-dimensional free energy surface at ($x_{(1,1)}$, $x_{(1,2)}$, $x_2 = 0.19$) (e), ($x_{(1,1)}$, $x_{(1,2)}$, $x_2 = 0.57$) (f), and ($x_{(1,1)}$, $x_{(1,2)}$, $x_2 = 1.0$) (g) with $N_{(1,1)} = 31$, $N_{(1,2)} = 64$, and $N_2 = 61$ (definition i). The cross-section at ($x_{(1,1)}$, $x_{(1,2)}$, $x_2 = 0.19$) (h), ($x_{(1,1)}$, $x_{(1,2)}$, $x_2 = 0.57$) (i), and ($x_{(1,1)}$, $x_{(1,2)}$, $x_2 = 1.0$) (j) with $N_{(1,1)} = 31$, $N_{(1,2)} = 64$, and $N_2 = 61$ (definition ii). The cross-section at ($x_1 = 0.19$, $x_{(2,1)}$, $x_{(2,2)}$) (k), ($x_1 = 0.47$, $x_{(2,1)}$, $x_{(2,2)}$) (l), ($x_1 = 0.68$, $x_{(2,1)}$, $x_{(2,2)}$) (m), and ($x_1 = 0.97$, $x_{(2,1)}$, $x_{(2,2)}$) (n) with $N_1 = 95$, $N_{(2,1)} = 31$, and $N_{(2,2)} = 30$ (definition iii). In (e-n) contour lines are drawn in every $2k_BT$. $\varepsilon/k_BT = 0.7$.

**Ankyrin repeat protein**

Fig.20 shows examples of free energy surface calculated for a 156-residue ankyrin repeat protein (PDB ID: 1mj0), which consists of repeated five helix-turn-helix sub-domains sd1, sd2, sd3, sd4, and sd5 (40). The two-dimensional representation of this repeat protein is constructed by dividing the chain into region 1 with $N_1 = 95$ for the N-terminal three sub-domains and region 2 with $N_2 = 61$ for the C-terminal two sub-domains. In the two-dimensional free energy surface of Fig. 20b, there are eight local minima corresponding to intermediate states, in addition to a minimum of the unfolded state and the global minimum at the native state of $x_1 \approx x_2 \approx 1.0$. There exist a number of multiple folding pathways connecting these intermediates.

In order to analyze these multiple folding pathways, we define the three-dimensional surface by using three different ways of dividing the protein chain: (i) The chain is divided into region$(1,1)_{\mathrm{i}}$, region$(1,2)_{\mathrm{i}}$, and region 2, where region$(1,1)_{\mathrm{i}}$ corresponds to sd1 with $N_{(1,1)} = 31$, and region$(1,2)_{\mathrm{i}}$ consists of sd2 and sd3 with $N_{(1,2)} = 64$. Region 2 consists of sd4 and sd5 with $N_2 = 61$. (ii) The chain is divided into region$(1,1)_{\mathrm{ii}}$, region$(1,2)_{\mathrm{ii}}$, and region 2, where region$(1,1)_{\mathrm{ii}}$ consists of sd1 and sd2 with $N_{(1,1)} = 64$, and region$(1,2)_{\mathrm{ii}}$ corresponds to sd3 with $N_{(1,2)} = 31$. (iii) The chain is divided into region 1, region(2,1), and region(2,2), where region 1 consists of sd1, sd2, and sd3 with $N_1 = 95$, region(2,1) corresponds to sd4 with $N_{(2,1)} = 31$, and region(2,2) corresponds to sd5 with $N_{(2,2)} = 30$.

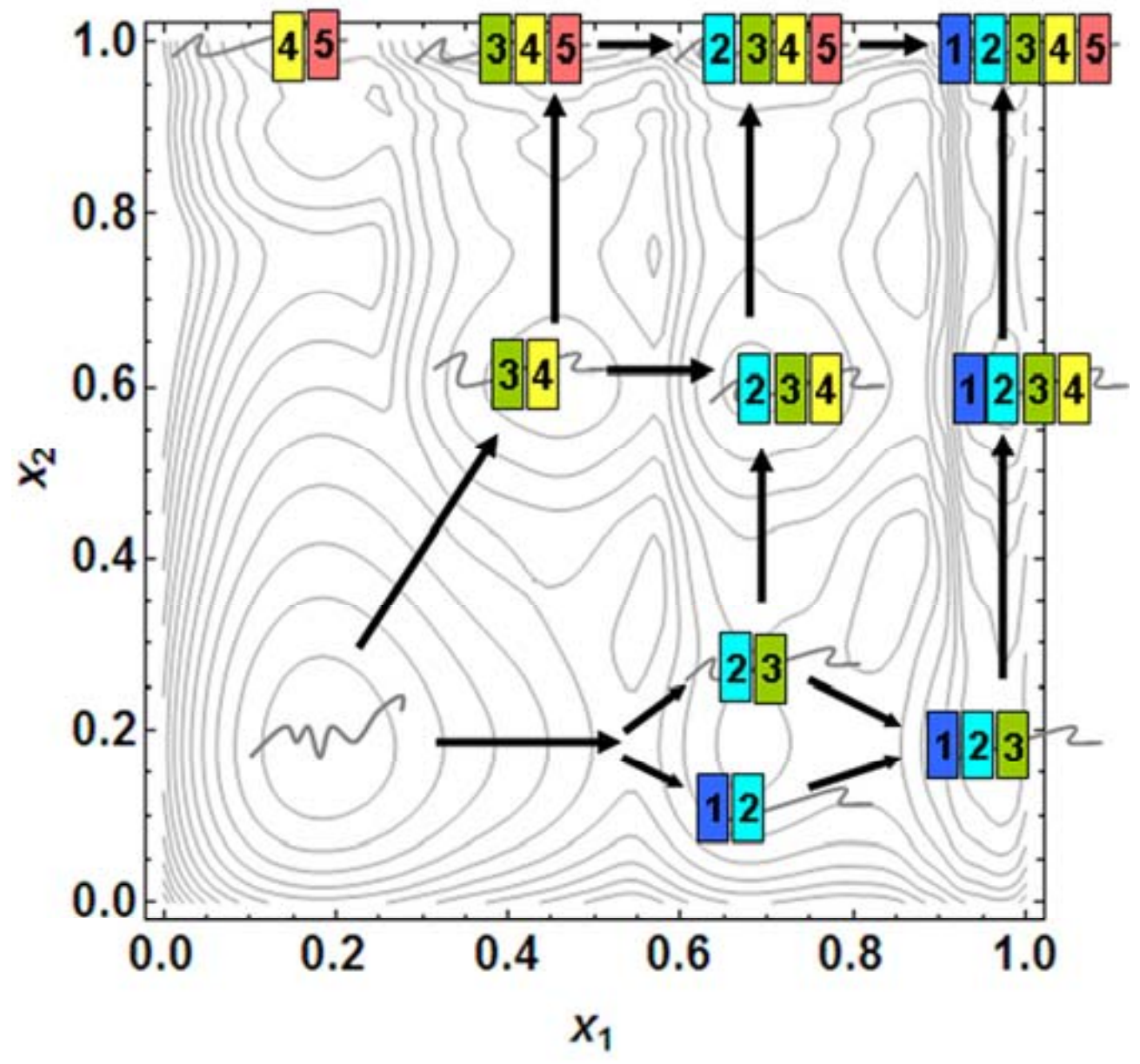

**FIGURE 21** Multiple folding pathways of the ankyrin repeat protein. The dominant folding pathways are denoted by real arrows, which pass through the free energy barriers with almost the same height.

Two-dimensional cross sections, ($x_{(1,1)}$, $x_{(1,2)}$, $x_2 = 0.19$), ($x_{(1,1)}$, $x_{(1,2)}$, $x_2 = 0.57$), and ($x_{(1,1)}$, $x_{(1,2)}$, $x_2 = 1.0$) of the three-dimensional surface are shown in Figs. 20e, 20f, and 20g for definition (i), and in Figs. 20h, 20i, and 20j for definition (ii), respectively. In Figs.20e and 20h, one can see that two dominant pathways coexist for folding of sd1-sd2-sd3 when sd4 and sd5 are kept unstructured at $x_2 = 0.19$. The free energy surfaces of Figs.20f, 20g, 20i, and 20j show that folding of sd1-sd2-sd3 is the three-state folding when sd4 and sd5 are partially structured at $x_2 = 0.57$ or well structured at $x_2 = 1.0$. In such three-state folding, sd2 and sd3 fold first and sd1 follows. Two-dimensional cross sections of three-dimensional surfaces with definition (iii), ($x_1 = 0.19$, $x_{(2,1)}$, $x_{(2,2)}$), ($x_1 = 0.47$, $x_{(2,1)}$, $x_{(2,2)}$), ($x_1 = 0.68$, $x_{(2,1)}$, $x_{(2,2)}$), and ($x_1 = 0.97$, $x_{(2,1)}$, $x_{(2,2)}$), are shown in Figs.20k, 20l, 20m and 20n. From these cross sections, we can find that sd4 folds first and sd5 follows along the dominant folding pathway. From these ten two-dimensional cross sections of three-dimensional surfaces, structural features and relative weight of multiple folding pathways can be clarified, which are summarized in Fig.21. Ferreiro et al. calculated the free energy surface of the same five-repeat protein and predicted that a dominant folding pathway is the one ongoing from N to C termini (41). In addition to this route, our results of multi-dimensional expressions predict the coexistence of multiple other pathways, which pass through free energy barriers with almost the same height.

**Higher dimensional surfaces**

In this way, various higher dimensional surfaces can be defined and their cross sections can be visualized in two-dimensional representation. In Figs.18-20, only a few slices of cross sections are shown for each three-dimensional surface, but the whole three-dimensional picture is obtained if we would add many slices of cross sections by using an efficient 3D graphics technique. The higher dimensional representation facilitates to scrutinize the free energy surface in a hierarchical fashion, which provides finer information on pathways and transition states. We expect that such fine information should be indeed necessary to explain important features of the experimental data.

## VII. SUMMARY AND DISCUSSIONS

In this paper the multi-dimensional representation of free energy surface was applied to several proteins, which exemplified usefulness of the method to investigate important thermodynamic and kinetic features of protein folding. Questions are more directly addressed with the multi-dimensional representation than with the one-dimensional representation on which region of protein folds faster than the other regions, how the structure is formed in transition states, whether there are intermediate states or not, or whether there are multiple parallel pathways. Features of structural fluctuations in each of unfolded, intermediate, and folded states are also better characterized on the multi-dimensional space than on a single reaction coordinate.

Further useful information on the mechanism of protein folding was obtained by decomposing the multi-dimensional free energy into several terms as in Eqs.25 and 26. For example, the functional form of $F_{\mathrm{e}}^{(i)}(x_i)$ determines whether the $i$th region is cooperative enough to fold in a two-state manner. The amplitude of $|F_{\mathrm{e}}^{(i)}(x_i)|$ is determined by the number of native interactions in the $i$th region, which largely affects the route of folding of the whole chain. The amplitude of $|U(\mathbf{x})|$ is determined by the number of native pairs interacting across the boundary of regions, which largely determines whether the multiple regions fold concurrently or there are pathways along which one of regions folds faster than the other. Asymmetry of $U(\mathbf{x})$ is also an important factor to determine the folding route. Since the functional form of $-TS_{\mathrm{e}}(\mathbf{x})$ largely affects the functional form of $F_{\mathrm{e}}^{(i)}(x_i)$, $F(\mathbf{x})$ tend to have a saddle point in the $L$-dimensional space at which $-TS_{\mathrm{e}}(\mathbf{x})$ shows a rapid change. These aspects of the multi-dimensional representation were shown in a clear way in a phenomenological model and were verified in example proteins by using a structure-based Hamiltonian model.

Among those variables, suitable quantities to examine how the protein chain can be divided are $F_{0i}(x_i)$, $F_{\mathrm{e}}^{(i)}(x_i)$, $U(\mathbf{x})$, and $-TS_{\mathrm{c}}^{(i)}(x_i)$. As can be seen from Eqs.25 and 26, those quantities satisfy the simple relations,

$$\begin{aligned} F(\mathbf{x}) &= \sum_{i=1}^{L} F_{0i}(x_i) + U(\mathbf{x}) \\ &= \sum_{i=1}^{L} \left[ F_{\mathrm{e}}^{(i)}(x_i) - TS_{\mathrm{c}}^{(i)}(x_i) \right] + U(\mathbf{x}), \end{aligned}$$

which is in contrast to the more complex relations, Eqs.24 and 25, satisfied by $E(\mathbf{x})$ or $-TS_{\mathrm{e}}(\mathbf{x})$. The divided individual regions should have a funnel-like feature by themselves when the condition of minimal frustration (44) or consistency among local and global structures (43) is well satisfied in protein as is realized in Go-type Hamiltonians. When individual multiple regions are funnel-like, each region can be further divided into smaller regions which should still have a funnel like feature. Such hierarchical repeat of division is illustrated in Fig.22. See APPENDIX C for more details on this hierarchical division. Decomposition of the funnel of repeat proteins was illustrated in a similar way (42), where the divided sub-domain should certainly have a funnel-like feature due to the specific one-dimensional structure of repeat proteins. In generic globular proteins, as shown in Fig.22, the funnel-like features of individual regions become weaker as the chain is divided into finer scales because the number of native interactions within individual regions is smaller and the number of native interactions across the different regions is larger as the size of each region becomes smaller. Through quantities such as $F_{0i}(x_i)$, $F_{\mathrm{e}}^{(i)}(x_i)$, $U(\mathbf{x})$, and $-TS_{\mathrm{c}}^{(i)}(x_i)$, the multi-dimensional theory provides insights on how the cooperativity or the funnel-like features within each region are lost or preserved as the division proceeds. Through the multi-dimensional analyses, we can tune resolution of theoretical "microscope" by choosing the way of dividing the chain.

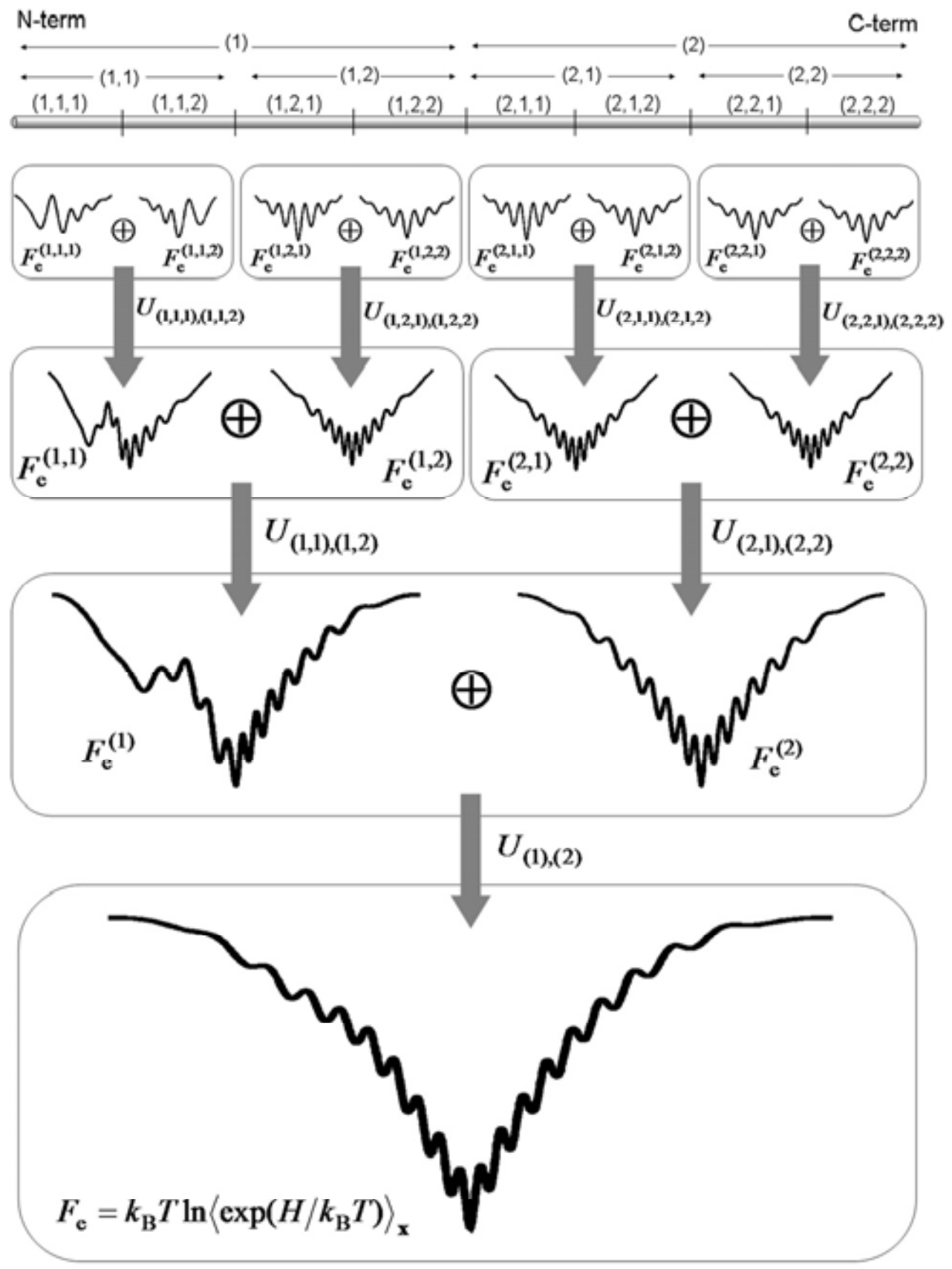

**FIGURE 21** Hierarchical decomposition of surface of effective energy $F_e$. = $k_B T\ln\langle\exp(\beta H)\rangle$. As the chain is divided from the entire protein chain to region $i$, to region $(i, j)$, and to region $(i, j, k)$ with $i, j, k = 1$ or 2, $F_e$ is decomposed into $F_e^{(i)}$, into $F_e^{(i,j)}$, and into $F_e^{(i,j,k)}$, which express cooperativity and funnel-like features of divided regions. Free energy due to interactions across regions are hierarchically decomposed into $U_{(1),(2)}$, into $U_{(i,j),(k,l)}$, and into $U_{(i,j,m),(k,l,n)}$, which express cooperativity among multiple regions. The symbol ⊕ represents synthesis of the decomposed free energy terms: When $F_e^{(i,j,1)}$ and $F_e^{(i,j,2)}$ are one-dimensional functions of $x_{(i,j,1)}$ and $x_{(i,j,2)}$, respectively, the synthesized effective energy, $F_e^{(i,j)}(x_{(i,j,1)}, x_{(i,j,2)})$, is calculated as $F_e^{(i,j)}(x_{(i,j,1)}, x_{(i,j,2)}) = F_e^{(i,j,1)}(x_{(i,j,1)}) + F_e^{(i,j,2)}(x_{(i,j,2)}) + U_{(i,j,1),(i,j,2)}(x_{(i,j,1)}, x_{(i,j,2)})$. Here, $x_{(i,j,1)}$ and $x_{(i,j,2)}$ are structural order parameters of region $(i,j,1)$ and region $(i,j,2)$, respectively. The same synthesis can be calculated at every stage in this figure. See APPENDIX C for more details. Folding scenario of the entire chain is classified in terms of cooperativity in each region and cooperativity among multiple regions. To choose the way of dividing the chain is to focus the theoretical "microscope" on the appropriate scale to resolve folding pathway.

Though we used the structure-based Hamiltonian model of Eq.30 to study example proteins, conclusions in the present paper should not be restricted to this particular choice of Hamiltonian model, but the framework of the multi-dimensional representation is applicable to other variants of models. Especially, Eqs.1-7, Eqs.19-26 and APPENDIX C explain model independent relations and are applicable to other Hamiltonian models. As natural extensions of the present model, we can adopt interactions with the site-dependent strength, or we can use the site-dependent entropic factor of $k_{\mathrm{B}}\ln\nu$. Use of off-lattice Go-model, instead of Eq.30, is most straightforward, and also by going beyond the Go-type models, it should be meaningful to analyze the effects of nonnative interactions with the multi-dimensional representation.

The multi-dimensional representation of free energy surface provides information on which region exhibits large fluctuation in native state, and how different regions correlate in structural fluctuation. Since these features may play roles in functional process especially in allosteric transformations, it is a challenging problem to apply the multi-dimensional method in analyses of protein functioning.

**ACKNOWLEDGMENTS**
This work was supported by Grants-in-aid for Scientific Research from the Ministry of Education, Culture, Sports, Science, and Technology, Japan, and from Japan Society for the Promotion of Science, and by grants for the 21st century COE program for Frontiers of Computational Science.

## APPENDIX A: CONTRIBUTIONS OF THE RUGGEDNESS IN THE SECOND-ORDER CUMULANT APPROXIMATION OF $k_B T \ln\langle\exp(\beta H_{nn})\rangle_{\mathbf{x}}$

Eq.10 is extended to general multi-dimensional case as

$$F_{\mathrm{e}}(\mathbf{x}) = k_{\mathrm{B}}T\ln\left\langle\exp(\beta H_{\mathrm{n}})\right\rangle_{\mathbf{x}}^{H_{\mathrm{n}}} + k_{\mathrm{B}}T\ln\left\langle\exp(\beta H_{\mathrm{nn}})\right\rangle_{\mathbf{x}}, \qquad (34)$$

from which we can see that the contribution of ruggedness to the free energy is

$$\begin{aligned} F_{\mathrm{rugg}}(\mathbf{x},T) &= k_{\mathrm{B}}T\ln\left\langle\exp(\beta H_{\mathrm{nn}})\right\rangle_{\mathbf{x}} \\ &= \sum_{n=1}^{\infty}\frac{\beta^{n-1}}{n!}C_n(H_{\mathrm{nn}};\mathbf{x}) \end{aligned}, \qquad (35)$$

where $C_n(H_{nn};\mathbf{x})$ is the $n$th order cumulant of $H_{nn}$ averaged with $H$.

Here, we replace the Hamiltonian $H = H_n + H_{nn}$ by $H(\eta) = H_n + \eta H_{nn}$. Then, the $n$th order cumulant of $H_{nn}$ averaged with $H(\eta)$, $C_n(H_{nn};\eta, \mathbf{x})$, is obtained from $C_{n-1}(H_{nn};\eta, \mathbf{x})$ by using the relation

$$C_n(H_{\mathrm{nn}};\eta,\mathbf{x}) = -k_{\mathrm{B}}T\frac{\partial C_{n-1}(H_{\mathrm{nn}};\eta,\mathbf{x})}{\partial\eta}. \qquad (36)$$

When $H_{nn}$ is treated as a Gaussian random variable as was assumed in the random energy model of folding (44), $C_n(H_{nn};\eta, \mathbf{x}) = 0$ for $n \geq 3$. In this case, the relation $\partial C_2(H_{nn};\eta, \mathbf{x})/\partial\eta = -\beta C_3(H_{nn};\eta, \mathbf{x}) = 0$ holds, so that the 2nd order of cumulant is written by an $\eta$-independent function $C_2(H_{nn};\eta, \mathbf{x}) = \Delta E^2(\mathbf{x}, T)$. Then, the first order of cumulant $C_1(H_{nn};\eta, \mathbf{x})$ is given by a linear function of $\eta$ as

$$\begin{aligned} C_1(H_{\mathrm{nn}};\eta,\mathbf{x}) &= -\frac{1}{k_{\mathrm{B}}T}\int_0^{\eta} d\eta'\Delta E^2(\mathbf{x},T) \\ &= -\frac{\Delta E^2(\mathbf{x},T)}{k_{\mathrm{B}}T}\eta\ . \end{aligned} \qquad (37)$$

Contribution of ruggedness for the Hamiltonian $H(\eta = 1)$ to the free energy $F_{\mathrm{rugg}}$ is

$$\begin{aligned} F_{\mathrm{rugg}}(\mathbf{x},T) &= C_1(H_{\mathrm{nn}};\eta=1,\mathbf{x}) + \frac{1}{2k_{\mathrm{B}}T}C_2(H_{\mathrm{nn}};\eta=1,\mathbf{x}) \\ &= -\frac{1}{2k_{\mathrm{B}}T}\Delta E^2(\mathbf{x},T) \end{aligned}. \qquad (38)$$

Then, contribution of ruggedness to the energy $E_{\mathrm{rugg}}(\mathbf{x},\mathrm{T}) = \partial(\beta F_{\mathrm{rugg}})/\partial\beta$ is obtained as

$$E_{\mathrm{rugg}}(\mathbf{x},T) = -\frac{1}{k_{\mathrm{B}}T}\Delta E^2(\mathbf{x},T) + \frac{1}{2k_{\mathrm{B}}}\frac{\partial\Delta E^2(\mathbf{x},T)}{\partial T}. \qquad (39)$$

## APPENDIX B: THE CONSTRAINED PARTITION FUNCTION FOR HAMILTONIAN OF EQ. 30

The partition function at the constraint $\mathbf{x} = (n_1, n_2, \cdots, n_L)$ in the $L$-dimensional coordinate is calculated from the generating function:

$$Q(\boldsymbol{\lambda}) \equiv \sum_{\mathbf{x}} Z(\mathbf{x}) \prod_{d=1}^{L} \lambda_d^{x_d}, \tag{40}$$

with $\boldsymbol{\lambda} = (\lambda_1, \lambda_2, \cdots, \lambda_L)$ as

$$Z(n_1, n_2, \cdots, n_L) = \prod_{d=1}^{L} \frac{1}{n_d!} \frac{\partial^{n_d}}{\partial \lambda_d^{n_d}} Q(\boldsymbol{\lambda}) \bigg|_{\boldsymbol{\lambda}=\mathbf{0}}. \tag{41}$$

$Q(\boldsymbol{\lambda})$ of Hamiltonian of Eq.30 can be calculated exactly by using the following recursive procedures (56): $Q(\boldsymbol{\lambda})$ is expressed as

$$Q(\boldsymbol{\lambda}) = P_{N-1}(1; \boldsymbol{\lambda}) \Omega(0). \tag{42}$$

$P_{N-1}(1; \boldsymbol{\lambda})$ is obtained by calculating the following recurrence equations:

$$P_0(1; \boldsymbol{\lambda}) = 1, \tag{43a}$$

$$P_0(l; \boldsymbol{\lambda}) = w_{N,N-l+2}(\boldsymbol{\lambda}), \ \ l \neq 1, \tag{43b}$$

$$P_k(1; \boldsymbol{\lambda}) = P_{k-1}(1; \boldsymbol{\lambda}) + P_{k-1}(2; \boldsymbol{\lambda}), \tag{43c}$$

$$P_k(l; \boldsymbol{\lambda}) = w_{N-k,N-k-l+2}(\boldsymbol{\lambda}) P_{k-1}(l; \boldsymbol{\lambda}) + P_{k-1}(l+1; \boldsymbol{\lambda}), \tag{43d}$$

$$l = 1,2 \cdots N-k+1, \ \ k = 1,2 \cdots N-1,$$

with

$$w_{j,i}(\boldsymbol{\lambda}) = \exp\Big[\sum_{r=i}^{j-1} \sum_{s=r+1}^{j} (\varepsilon_{r,s} / k_{\mathrm{B}} T) \Delta_{r,s} - \sum_{r=i}^{j} \ln \nu_r\Big] \prod_{d=1}^{L} \lambda_d^{X_d(j,i)}. \tag{44}$$

$X_k(j,i)$ is defined as

$$X_d(j,i) = \sum_{k=i}^{j} \eta_d(k), \tag{45}$$

where $\eta_d(k) = 1$ when the $k$th residue is in the $d$th region, $\eta_d(k) = 0$ otherwise.

## APPENDIX C: HIERARCHICAL DECOMPOSITION OF FREE ENERGY SURFACE

We first divide protein into two regions. This is the $L$=2 case of Eq.21, and we have

$$F_{\mathrm{e}}(x_1, x_2) = F_{\mathrm{e}}^{(1)}(x_1) + F_{\mathrm{e}}^{(2)}(x_2) + U(x_1, x_2). \tag{46}$$

In the next step, each region of $i$ = 1, 2 is further divided into two regions, ($i$,1) and ($i$,2). Then, $F_{\rm e}^{(i)}(x_i)$ in Eq.46 is decomposed as

$$F_{\rm e}^{(i)}(x_{(i,1)},x_{(i,2)})=F_{\rm e}^{(i,1)}(x_{(i,1)})+F_{\rm e}^{(i,2)}(x_{(i,2)})+U_{(i,1),(i,2)}(x_{(i,1)},x_{(i,2)}), \quad (47)$$

where $F_{\rm e}^{(i,j)}(x_{(i,j)})$ with $j$ = 1, 2 is the effective energy of region ($i$, $j$) obtained by subtracting the conformational entropy from free energy and $U_{(i,1),(i,2)}(x_{(i,1)},x_{(i,2)})$ is the free energy of interactions between regions ($i$,1) and ($i$,2). This division can be successively continued as in Fig.22. At the $k$th division, we have $L$=$2^k$ regions of $(i_1,i_2,\cdots,i_k)$. Region $(i_1,i_2,\cdots,i_k)$ is the $j$th region from the N-terminus with $j=\sum_{l=1}^{k}(i_l-1)2^{k-l}+1$. When this region is further divided into $(i_1,\cdots,i_k,1)$ and $(i_1,\cdots,i_k,2)$, the effective energy is decomposed as

$$\begin{aligned}F_{\rm e}^{(i_1,\cdots,i_k)}(x_{(i_1,\cdots,i_k,1)},x_{(i_1,\cdots,i_k,2)})&=F_{\rm e}^{(i_1,\cdots,i_k,1)}(x_{(i_1,\cdots,i_k,1)})\\&+F_{\rm e}^{(i_1,\cdots,i_k,2)}(x_{(i_1,\cdots,i_k,2)})\\&+U_{(i_1,\cdots,i_k,1),(i_1,\cdots,i_k,2)}(x_{(i_1,\cdots,i_k,1)},x_{(i_1,\cdots,i_k,2)}),\end{aligned} \quad (48)$$

At the $K$th division, we have $L$=$2^K$ regions. Using the recurrent relation of Eq.48, we have

$$F(\mathbf{\Gamma})=F_0(\mathbf{x})+U(\mathbf{\Gamma}), \quad (49)$$

$$F_0(\mathbf{x})=\sum_{i_1=1}^{2}\sum_{i_2=1}^{2}\cdots\sum_{i_K=1}^{2}F_{\rm e}^{(i_1,i_2,\cdots,i_K)}(x_{(i_1,\cdots,i_K)})-TS_{\rm c}(\mathbf{x}), \quad (50)$$

$$S_{\rm c}(\mathbf{x})=\sum_{i_1=1}^{2}\sum_{i_2=1}^{2}\cdots\sum_{i_K=1}^{2}S_{\rm c}^{(i_1,i_2,\cdots,i_K)}(x_{(i_1,\cdots,i_K)})+k_{\rm B}\ln\Omega(\mathbf{1}), \quad (51)$$

$$\begin{aligned}U(\mathbf{\Gamma})=&U_{(1),(2)}(x_{(1)},x_{(2)})+\sum_{i_1}U_{(i_1,1),(i_1,2)}(x_{(i_1,1)},x_{(i_1,2)})+\cdots\\&+\sum_{i_1,\cdots,i_{K-1}}U_{(i_1,\cdots,i_{K-1},1),(i_1,\cdots,i_{K-1},2)}(x_{(i_1,\cdots,i_{K-1},1)},x_{(i_1,\cdots,i_{K-1},2)}),\end{aligned} \quad (52)$$

where $\mathbf{\Gamma}$ is a set of multi-dimensional coordinates defined by

$$\mathbf{\Gamma}(K)=\{x_{(i_1,i_2,\cdots,i_l,\cdots,i_k)}\,|\,i_l=1,2,\quad l=1,\cdots,k,\quad k=1,\cdots,K\},$$

$\mathbf{x}\subset\mathbf{\Gamma}$, and $U(x_1,x_2)$ in Text is here rewritten by $U_{(1),(2)}(x_{(1)},x_{(2)})$. The $2^{K-1}+r$ dimensional expression of the free energy with $r$ = 1,2,…, $2^{K-1}-1$ can be written as

$$F(\mathbf{\Lambda}(\mathbf{p}))=F_0(\mathbf{\Lambda}'(\mathbf{p}))+U(\mathbf{\Lambda}(\mathbf{p})), \quad (53)$$

$$
\begin{aligned}
F_0(\boldsymbol{\Lambda}'(\mathbf{p})) = &\sum_{\mathbf{p}'(r)} F_{\mathrm{e}}^{(i_1,\cdots,i_{K-1})}(x_{(i_1,\cdots,i_{K-1})}) \\
&+ \sum_{\mathbf{p}(2^K-r)} [F_{\mathrm{e}}^{(i_1,\cdots,i_{K-1},1)}(x_{(i_1,\cdots,i_K,1)}) + F_{\mathrm{e}}^{(i_1,\cdots,i_{K-1},2)}(x_{(i_1,\cdots,i_K,2)})] \\
&- TS_{\mathrm{c}}(\mathbf{x}) \quad ,
\end{aligned} \tag{58}
$$

$$
\begin{aligned}
U(\boldsymbol{\Lambda}(\mathbf{p})) = &U_{(1),(2)}(x_{(1)}, x_{(2)}) + \sum_{i_1} U_{(i_1,1),(i_1,2)}(x_{(i_1,1)}, x_{(i_1,2)}) + \cdots \\
&+ \sum_{\mathbf{p}(2^K-r)} U_{(i_1,\cdots,i_{K-1},1),(i_1,\cdots,i_{K-1},2)}(x_{(i_1,\cdots,i_{K-1},1)}, x_{(i_1,\cdots,i_{K-1},2)}) \quad ,
\end{aligned} \tag{59}
$$

where $\mathbf{p}'(r)$ denotes a set of $r$ elements chosen from the set $\{(i_1,i_2,\ldots,i_{K-1})\}$ and $\mathbf{p}(2^K-r)$ is a set of the rest $2^{K-1} - r$ elements. $\boldsymbol{\Lambda}(\mathbf{p})$ and $\boldsymbol{\Lambda}'(\mathbf{p}')$ are sets of multi-dimensional coordinates defined by

$$
\boldsymbol{\Lambda}(\mathbf{p}) = \boldsymbol{\Gamma}(K-1) \cup \{x_{(i_1,\cdots,i_{K-1},j)} \mid j = 1,2, (i_1,\cdots,i_{K-1}) \in \mathbf{p}\} ,
$$

and

$$
\boldsymbol{\Lambda}'(\mathbf{p}) = \{x_{(i_1,\cdots,i_{K-1})} \mid (i_1,\cdots,i_{K-1}) \in \mathbf{p}'\} \cup \{x_{(i_1,\cdots,i_{K-1},j)} \mid j = 1,2, (i_1,\cdots,i_{K-1}) \in \mathbf{p}\} .
$$

## REFERENCES

1. A. R. Fersht, *Structure and Mechanism in Protein Science: A Guide to Enzyme Catalysis and Protein Folding* (Freeman & Company, New York) (1999).
2. C. M. Dobson, A. Sali, and M. Karplus, *Angew. Chem. Int. Ed. Engl.* **37,** 868 (1998).
3. J. N. Onuchic, Z. Luthey-Schulten , and W. G. Wolynes, *Ann. Rev. Phys. Chem*. **48**, 545 (1997).
4. J. N. Onuchic and P. G. Wolynes, *Curr. Opn. Struct. Biol.* **14**, 70 (2004).
5. S. S. Cho, Y. Levy, and P. G. Wolynes, *Proc. Natl. Acad. Sci. USA* **103**, 586 (2006).
6. D. J. Brockwell and S. E. Radford, *Curr. Opn. Struct. Biol.* **17**, 30 (2007).
7. K. Itoh and M. Sasai, *Proc. Natl. Acad. Sci. USA* **103**, 7298 (2006).
8. M. O. Lindberg and M. Oliveberg, *Curr. Opn. Struct. Biol.* **17**, 21 (2007).
9. D. U. Ferreiro, A. M. Walczak, E. A. Komives, and P. G. Wolynes, *PLoS Comput. Biol.* **16**, e1000070 (2008).
10. K. Itoh and M. Sasai, *Proc. Natl. Acad. Sci. USA*, **105**, 13865 (2008).
11. Y. Levy, P. G. Wolynes, and J. N. Onuchic, *Proc. Natl. Acad. Sci. USA* **101**, 511 (2004).
12. Y. Levy, S. S. Cho, J. N. Onuchic, and P. G. Wolynes, *J. Mol. Biol.* **346**, 1121 (2005).
13. N. D. Socci, J. N. Onuchic and P. G. Wolynes, *Proteins* **32**, 136 (1998).
14. F. B. Sheinerman and C. L. Brooks II, *J. Mol. Biol*. **278**, 439 (1998).
15. D. E. Otzen, O. Kristensen, M. Proctor, and M. Oliveberg, *Biochemistry* **38**, 6499 (1999).
16. D. E. Otzen and M. Oliveberg, *J. Mol. Biol.* **317**, 613 (2002).
17. I. A. Hubner, M. Oliveberg, and E. I. Shakhnovich, *Proc. Natl. Acad. Sci. USA* **101**, 8354 (2004).
18. M. Olofsson, S. Hansson, L. Hedberg, D. T. Logan, and M, Oliveberg, *J. Mol. Biol.* **365**, 237 (2007).
19. M. Oliveberg, *Curr. Opn. Struct. Biol.* **11**, 94 (2001).
20. V. P. Grantcharova and D. Baker., *Biochemistry* **36**, 15685 (1997).
21. V. P. Grantcharova, D. S. Riddle, J. V. Santiago, and D. Baker, *Nat. Struct. Boil.* **5**, 714 (1998).
22. A. R. Lam, J. M. Borreguero, F. Ding, N.V. Dokholyan, S. V. Buldyrev, H. E. Stanley, and E. Shakhnovich, *J. Mol. Biol.* **373**, 1348 (2007).
23. E. López-Hernández and L. Serrano, *Fold. Des.* **1**, 43 (1996).
24. P. Garcia,. L. Serrano, M. Rico, and M. Bruix, *Structure* **10**, 1173 (2002).

25. Y. J. M. Bollen, C. P. M. van Mierlo, *Biophys. Chem.* **114**, 181 (2005).
26. A. R. Fersht, *Proc. Natl. Acad. Sci. USA* **97**, 14121 (2000).
27. L. Serrano, A. Matouschek, and A. R. Fersht, *J. Mol. Biol.* **224**, 805 (1992).
28. X. Salvatella, C. M. Dobson, A. R. Fersht, and M. Vendruscolo, *Proc. Natl. Acad. Sci. USA* **102**, 12389 (2005).
29. M. M. Garcia-Mira, M Sadqi, N. Fischer, J. M. Sanchez-Ruiz, V. Muñoz, Science **298**, 2191 (2002).
30. M. Sadqi, D. Fushman, and V. Munoz, *Nature* **442**, 317 (2006).
31. N. Ferguson, P. J. Schartau, T. D. Sharpe, S. Sato, and A. R. Fersht, *J. Mol. Biol.* **344**, 295 (2004).
32. N. Ferguson, T. D. Sharpe, P. J. Schartau, S. Sato, M. D. Allen, C. M. Johnson, T. J. Rutherford, and A. R. Fersht, *J. Mol. Biol*. **353**, 427 (2005).
33. S. S. Cho, P. Weinkam, and P. G. Wolynes, *Proc. Natl. Acad. Sci. USA* **105**, 118 (2008).
34. N. Ferguson, T. D. Sharpe, C. M. Johnson, P. J. Schartau, and A. R. Fersht, *Nature* **445**, E14 (2007).
35. Z. Zhou and Y. Bai, *Nature* **445**, E16 (2007).
36. M. Sadqi, D. Fushman, V. Muñoz, *Nature* **445**, E17 (2007).
37. W. Yu, K. Chung, M. Cheon, M. Heo, K-H. Han, S. Ham, and I. Chang, *Proc. Natl. Acad. Sci. USA* **105**, 2397 (2008).
38. D. Kern, B. F. Volkman, P. Luginbuhl, M. J. Nohaile, S. Kustu, and D. E. Wemmer, *Nature* **402**, 894 (1999).
39. B. F. Volkman, D. Lipson, D. E. Wemmer, and D. Kern, *Science* **291**, 2429 (2001).
40. A. Kohl, H. K. Binz, P. Forrer, M. T. Stumpp, A. Plückthun, and M. G. Grütter, *Proc. Natl. Acad. Sci. USA* **100**, 1700 (2003).
41. D. U. Ferreiro, S. S. Cho, E. A. Komives, and P. G. Wolynes, *J. Mol. Biol.* **354**, 679 (2005).
42. D. U. Ferreiro and P. G. Wolynes, *Proc. Natl. Acad. Sci. USA* **105**, 9853 (2008).
43. N. Go, *Annu. Rev. Biophys. Bioeng*. **12**, 183 (1983).
44. J. D. Bryngelson and P. G. Wolynes, *Proc. Natl. Acad. Sci. USA* **84**, **7524** (1987).
45. J. D. Bryngelson and P. G. Wolynes, *J. Phys. Chem.* **93**, 6902 (1989).
46. R. Zwanzig, *Proc. Natl. Acad. Sci. USA* **92**, 9801 (1995).
47. S. S. Plotkin and J. N. Onuchic, *Quart. Rev. Biophys*. **35**, 111 (2002).
48. S. S. Plotkin and J. N. Onuchic, *Quart. Rev. Biophys*. **35**, 205 (2002).
49. P. Das, S. Matysiak, and C. Clementi, *Proc. Natl. Acad. Sci. USA*, **102**, 10141 (2005).

50. S. S. Plotkin, J. Wang, and P. G. Wolynes, *J. Chem. Phys.* **106**, 2932 (1997).
51. H. Wako and N. Saito, *J. Phys. Soc. Jpn.* **44**, 1931 (1978).
52. H. Wako and N. Saito, *J. Phys. Soc. Jpn.* **44**, 1939 (1978).
53. V. Muñoz and W.A.Eaton, *Proc. Natl. Acad. Sci. USA* **96**,11311 (1999).
54. E. R. Henry and W. A. Eaton, *Chem. Phys*. **307**, 163 (2004).
55. N. Go and H. Abe, *Biopolymers* **20**, 991 (1981) .
56. P. Bruscolini and A. Pelizzola, *Phys. Rev. Lett.* **88**, 258101 (2002).
57. M. Zamparo and A. Pelizzola, *Phys. Rev. Lett.* **97**, 068106 (2006).
58. P. Bruscolini, A. Pelizzola, and M. Zamparo, *Phys. Rev. Lett.* **99**, 038103 (2007).
59. E. D. Nelson and N. V. Grishin, *Proc. Natl. Acad. Sci. USA* **105**, 1489 (2008).
60. H. S. Chung and A. Tokmakoff. *Proteins* **72**, 488 (2008).
61. A. Imparato, A. Pelizzola, and M. Zamparo, *Phys. Rev. Lett.* **98**, 148102 (2007).
62. A. Imparato, and A. Pelizzola, *Phys. Rev. Lett.* **100**, 158104 (2008).
63. K. Itoh and M. Sasai, *Proc. Natl. Acad. Sci. USA* **101**, 14736 (2004).
64. J. Karanicolas and C. L. Brooks, *Proteins Struct. Funct. Gen*. **53**, 740 (2003).
65. S. Huang, K. S. Ratliff, M. P. Schwartz, J. M. Spenner and A. Matouschek, *Nat. Struct. Biol.* **6**, 1132 (1999).